\def\IDvalue{KL}
\def\titlevalue{Indices for 6 dimensional superconformal field theories}

\def\authorvalue{Seok Kim$^1$ and Kimyeong Lee$^2$}
\def\shortauthorvalue{Seok Kim and Kimyeong Lee}

\def\addressvalue{
$^1$Department of Physics and Astronomy \& Center for
Theoretical Physics,\\
Seoul National University, Seoul 151-747, Korea\\
$^2$School of Physics, Korea Institute for Advanced Study,
Seoul 130-722, Korea\\
 {\tt skim@phya.snu.ac.kr, klee@kias.re.kr}}

\def\abstractvalue{We review some recent developments in the 6 dimensional $(2,0)$ superconformal
field theories, focusing on their BPS spectra in the Coulomb and symmetric phases
computed by various Witten indices. We shall discuss the instanton partition function
of 5d maximal super-Yang-Mills theory, and the 6d superconformal index.}

\def\preprintvalue{
 SNUTP15-001\\  KIAS-P15008}

\ifx\ifLONG\undefined
\documentclass[12pt]{article}
\newcommand{\chapterauthor}[1]{
\begin{center}
{\bf \normalsize  #1}
\end{center}
}

\newcommand{\chapteraddress}[1]{
\begin{center}
{ \small \it \addressvalue}
\end{center}
}

\newcommand{\chapterabstract}[1]{
\vspace{\baselineskip}
\begin{center}
\textbf{\small Abstract}
\end{center}
#1}

\newcommand{\chapterheader}{

\chapter[\titlevalue{}  (by \shortauthorvalue)]{\titlevalue}
\label{Chapter\IDvalue}
\chapterauthor{\authorvalue}
\chapteraddress{\addressvalue}
\chapterabstract{\abstractvalue}
\tightmtctrue
\minitoc
}

\newcommand{\documentheader}{
\begin{flushright} \small
  \preprintvalue
 \end{flushright}

\begin{center}
{\bf \Large \titlevalue}
\end{center}

\chapterauthor{\authorvalue}
\chapteraddress{\addressvalue}
\chapterabstract{\abstractvalue}

\medskip

This is a contribution to the review volume ``Localization techniques
in quantum field theories'' (eds. V.~Pestun and M.~Zabzine) which
contains 17 Chapters available at \cite{ContributionSummary}

\tableofcontents
}

\newcommand{\ifvolume}[2]{\ifx\ifLONG\undefined#2\else#1\fi}

\newcommand{\documentfinish}{
\ifx\ifLONG\undefined
\bibliographystyle{bibreview} 
\bibliography{\IDvalue,review}  
\end{document}
\else

\fi
}

\newcommand{\documentfinishBBL}{
\addcontentsline{toc}{section}{References}
\ifx\ifLONG\undefined

\end{document}
\else

\fi
}

\ifx\ifLONG\undefined

\else 

\fi

\newcommand{\ContributionSummaryBibItemReference}
{
\bibitem{ContributionSummary}
V.~Pestun and M.~Zabzine, eds., {\em Localization techniques in quantum field
  theory}, vol.~xx.
\newblock Journal of Physics A, 2016.
\newblock \href{http://arxiv.org/abs/1608.02952}{{\tt 1608.02952}}.
\newblock \url{https://arxiv.org/src/1608.02952/anc/LocQFT.pdf},
  \url{http://pestun.ihes.fr/pages/LocalizationReview/LocQFT.pdf}.
}

\usepackage{amsfonts,amssymb,epsfig,amsmath}
\usepackage{color}
\usepackage{fullpage}

\numberwithin{equation}{section}

\usepackage{hyperref}
\begin{document}
\thispagestyle{empty}
\documentheader
\else\chapterheader \fi

\section{Introduction}

With various string dualities found in mid 90's, interacting quantum field theories 
in spacetime dimensions larger than $4$ were
discovered from string theory \cite{Witten:1995zh,Seiberg:1996vs,Seiberg:1996bd}.
Many aspects of these QFTs are counterintuitive from the conventional viewpoint and have 
enriched our notion on
what quantum field theory is. The higher dimensional QFTs are also the key to understanding
the strong-coupling aspects of string and M theories. Multiple M5-branes and
6d $(2,0)$ theories are such examples.

However, we still do not know their intrinsic definitions. For instance, they are
strongly interacting CFTs, and no Lagrangian descriptions are known. Despite this situation, in the last few years there has been interesting progress in our understanding
on the 5 and 6 dimensional superconformal field theories, based on various effective
descriptions of these theories. In particular, we shall focus on the 
advances in supersymmetric observables of these higher dimensional field theories.

There have been many works on the BPS observables of 5d and 6d SCFTs,
especially from 2012 when the techniques of curved space SUSY QFT
were applied to higher dimensions. For instance, in 5d SCFTs, there have been
extensive studies on the partition functions on $S^4\times S^1$
\cite{Kim:2012gu,Bergman:2013koa,Hayashi:2013qwa,Bao:2013pwa} and $S^5$
\cite{Jafferis:2012iv,Alday:2014rxa,Assel:2012nf}. There have also been many studies on 
6d SCFTs. Their partition functions were studied on $S^5\times S^1$ \cite{Kim:2012ava,Kallen:2012zn,Kim:2012tr,Lockhart:2012vp,Kim:2012qf,
Minahan:2013jwa,Kim:2013nva}, $S^3\times S^1\times M_2$ \cite{Kawano:2012up},
$S^2\times S^1\times M_3$ \cite{Yagi:2013fda,Lee:2013ida}, and $S^3\times M_3$
\cite{Cordova:2013cea}, where $M_2$ and $M_3$ are 2 and 3 dimensional manifolds. Various 
6d defect partition functions on curved manifolds were also studied, such as the dimension 
$2$ surfaces \cite{Kim:2012qf,Minahan:2013jwa,Mori:2014tca,Bullimore:2014upa} and
dimension $4$ surfaces \cite{Nieri:2013yra,Bullimore:2014upa}. The progress was made possible 
largely due to the technical advances in 5d super-Yang-Mills theories on curved manifolds.
See \cite{Kallen:2012cs,Hosomichi:2012ek,Kallen:2012va,Kim:2012ava,Imamura:2012bm} and
references therein for some early developments, \cite{Cordova:2013bea} for some
systematic formulations on 5d maximal SYM on curved backgrounds,
\cite{Kim:2012gu,Lockhart:2012vp,Kim:2012qf,Nieri:2013vba,Qiu:2013aga,Qiu:2014oqa} for
the factorizations of 5d partition functions on $S^4\times S^1$ and $S^5$,
\cite{Qiu:2014cha} for the saddle point structures of the supersymmetric path integral
of 5d SYM on Sasaki-Einstein spaces. Often, via factorization, some curved space 
observables are related to those of the same QFT on flat spacetime, such as $\mathbb{R}^4\times S^1$ 
or $\mathbb{R}^4\times T^2$, in the Coulomb phase. The last Coulomb phase observables have been 
studied from relatively long time ago, after the pioneering works by Nekrasov et al.
\cite{Nekrasov:2002qd,Nekrasov:2003rj}. There have been continuing developments in
these observables \cite{Kim:2011mv,Haghighat:2013gba,Haghighat:2013tka,Haghighat:2014pva,
Hwang:2014uwa,Kim:2014dza,Cai:2014vka,Gaiotto:2014ina,Haghighat:2014vxa}, especially
in the recent few years after the realization of their relations to the conformal phase 
observables.

Especially  in this review paper, we shall discuss the BPS spectra of these theories 
captured by Witten index partition functions. The main objects will be the partition
functions of 6d SCFTs on the Omega deformed $\mathbb{R}^4\times T^2$ in the Coulomb phase,
and also the superconformal index partition function on $S^5\times S^1$. We shall mostly
discuss the $(2,0)$ CFTs, since major progress has been made only for these theories 
so far. We shall however comment on possible generalizations to a wider class of 
$(1,0)$ CFTs at various places. It will mostly be reviews of some papers cited 
above, but contains some unpublished materials as well. In the
rest of the introduction, we shall briefly motivate the objects that we study in this paper and also our methods and approaches.

One observable discussed in this paper is the superconformal index of the 6d
SCFT \cite{Bhattacharya:2008zy}. This is a Witten index which counts
BPS local operators of the CFT on $\mathbb{R}^6$. Or equivalently, it counts BPS states
of the radially quantized CFT on $S^5\times\mathbb{R}$, weighted by various chemical
potentials. Being a supersymmetric version of the thermal partition function, we
can regard it as the partition function on $S^5\times S^1$ with supersymmetric boundary
conditions of fields along $S^1$. Schematically, we shall be considering expressions for
this index of the form
\begin{equation}\label{KL6d-index-factorize}
  Z_{S^5\times S^1}(\mu)=\int[d\phi]e^{-S_{0}(\phi)}
  Z_{\mathbb{R}^4\times T^2}^{(1)}(\phi,\mu)Z_{\mathbb{R}^4\times T^2}^{(2)}(\phi,\mu)
  Z_{\mathbb{R}^4\times T^2}^{(3)}(\phi,\mu)\ ,
\end{equation}
where $\phi$ denotes the `scalar VEV in the Coulomb branch' which is integrated over
in the above expression, and $S_0(\phi)$ is the so-called `classical action' which
shall be explained later. The three ingredients $Z_{\mathbb{R}^4\times T^2}^{(i)}(\phi,\mu)$ 
are the Coulomb branch Witten index of the circle comapactified 6d theory on flat space, 
which we shall explain in detail in section 2. $\mu$ collectively denotes the chemical potentials.
In particular, it will contain the (dimensionless) `inverse temperature' like variable
$\beta=\frac{2\pi r_1}{r_5}$,
where $r_1,r_5$ are the radii of the $S^1$ and $S^5$ factor, respectively.
Other chemical potentials, in our parametrization, will be the three rotation
chemical potentials $\omega_1,\omega_2,\omega_3$ on $S^5$, and
those for the flavor symmetries.

The expression above is just one of the many occasions in which the SUSY QFT partition
functions on compact manifolds are related to the Coulomb branch partition functions.
A canonical example can be found for gauge theories on $S^4$ \cite{Pestun:2007rz}, related
to the Coulomb phase partition function on $\mathbb{R}^4$. The Coulomb branch partition
function has been an extremely useful observable by itself, for many reasons, and has been
extensively studied since \cite{Nekrasov:2002qd,Nekrasov:2003rj}. In the context of 6d CFTs,
it provides useful information on the BPS spectrum of wrapped self-dual strings
\cite{Strominger:1995ac,Howe:1997ue}.
Also, understanding its properties better has been (and will be) the key to the developments
in the conformal phase observables, such as (\ref{KL6d-index-factorize}). So our
section 2 will review the old and new developments on the Coulomb
branch partition function on $\mathbb{R}^4\times T^2$. Somewhat interestingly, the recent
demand on refined understanding of this observable triggered a technically clearer
derivation of this rather old observable, especially for many subtle QFTs for which this
partition function could not be computed before.

Coming back to the superconformal index (\ref{KL6d-index-factorize}), we do not
have a self-contained formulation to justify it. However, considering the
regime with small circle,
$\beta\ll 1$, we can try to understand the structure of (\ref{KL6d-index-factorize})
using a 5 dimensional effective description. When $\beta\ll 1$, the expression
(\ref{KL6d-index-factorize}) admits a `weak coupling' expansion in $\beta$, either
perturbative one in power series of $\beta$, or nonperturbative one in a series of
$e^{2\pi i\tau_i}\ll 1$, where $\tau_i=\frac{2\pi i}{\beta\omega_i}$. The last 
`weak coupling' expansion acquires a more
precise sense when the 6d SCFT compactified on a small circle admits a weakly coupled
5d Yang-Mills theory description. For instance, when we compactify the 6d $(2,0)$ CFT
of ADE type on small $S^1$ with radius $r_1$, then at low energy we would have a 5
dimensional maximal super-Yang-Mills description on $S^5$.\footnote{`Maximal SYM'
will often mean a QFT with the field content of maximal SYM, subject to
deformations due to curvature and chemical potential parameters. So the number of
preserved SUSY could be less than $16$. For instance, mass-deformed maximal SYM, 
the $\mathcal{N}=2^\ast$ theory, will often be called just maximal SYM.}
Such a 5d SYM limit exists for some other $(1,0)$ SCFTs.\footnote{We shall comment on
cases in which no 5d SYM limits exist, in which case the expression (\ref{KL6d-index-factorize}) could 
still make sense.}
The radius $r_1$ of the circle gets mapped to the 5d gauge groupling $g_{YM}$ via
\begin{equation}
  \frac{4\pi^2}{g_{YM}^2}=\frac{1}{r_1}\ ,
\end{equation}
in our convention for $g_{YM}$. So here, the small $\beta$ expansion is indeed the
weak coupling expansion.

The partition function $Z^{(i)}_{\mathbb{R}^4\times T^2}$ at $\beta\ll 1$ thus reduces
to 5d SYM partition functions on $\mathbb{R}^4\times S^1$, which has been studied
in great detail since \cite{Nekrasov:2002qd}. This decomposes into the perturbative
part and instanton corrections,
\begin{equation}
  Z^{(i)}_{\mathbb{R}^4\times T^2}=Z_{\rm pert}^{(i)}(\phi,\omega,m)
  Z_{\rm inst}^{(i)}(\beta,\phi,\omega,m)\ ,
\end{equation}
where $Z_{\rm pert}^{(i)}$ is the 1-loop contribution which is independent of $\beta$,
and
\begin{equation}
  Z_{\rm inst}^{(i)}=\sum_{k=0}^\infty e^{-\frac{4\pi^2k}{\beta\omega_i}}
  Z_{k}^{(i)}(\phi,\omega,m)
\end{equation}
with $Z_0\equiv 1$ acquires contributions from Yang-Mills instantons localized
on $\mathbb{R}^4$ and extended along $S^1$. These instanton solitons in 5d SYM are
interpreted as Kaluza-Klein modes of the 6d CFT compactified on circle, so captures
nontrivial $\beta$ dependence even after compactification on small circle.
$S_0(\phi)$ in (\ref{KL6d-index-factorize}) can also be computed from 5d SYM. 
So pragmatically, we shall be able to understand all the ingredients of
(\ref{KL6d-index-factorize}) from 5d SYM.
Having obtaining the weakly coupled expression (\ref{KL6d-index-factorize}) for the
6d index, one may sum over the $k$ series and re-expand the result at $\beta\gg 1$
if one has a good technical control over $Z_{\rm inst}^{(i)}$. The strong coupling result
is useful because the spectral information can be obtained only after the expansion in
the small fugacity $e^{-\beta}\ll 1$. We explain in section 3.2 how
to explicitly do this in some special cases.

At this point, we also note that there is another version of the 6d index formula
taking the form (\ref{KL6d-index-factorize}), which is obtained from 5d SYM on
$\mathbb{CP}^2\times S^1$. This expression takes a manifest form of the index,
given as an expansion in $e^{-\beta}$ at $\beta\gg 1$. We shall explain it for the
$(2,0)$ theory in section 3.3, emphasizing its virtue and new physics visible from 
this setting.

Conceptually, it will be interesting to understand whether the formulae of the type 
(\ref{KL6d-index-factorize}) are correct for all 6d SCFTs without relying on 5d 
SYM descriptions. Also, it would be nice to understand whether it is the unexpected feature 
of 5d SYM or our
specific choice of SUSY observables which made 5d SYM useful here. For the
Coulomb phase index explained in section 2, we can completely bypass the 5d SYM description
logically (although it is still useful), and directly compute the index from string/M-theory
by taking decoupling limits and starting from UV complete 1d or 2d gauge theories.
We do not know whether we can bypass the 5d SYM description for the superconformal index.

The remaining part of this paper is organized as follows. In section 2, we explain
the computation and physics of the Coulomb branch indices of 6d CFTs on Omega deformed 
$\mathbb{R}^4\times T^2$, mainly from 1 dimensional gauge theories (also with detailed
comments on studies from 2d gauge theories). In section 3, we explain the 6d $(2,0)$ 
superconformal  index and the physics contained in it. Section 4 concludes with 
open questions and comments. Appendix A elaborates on the SUSY gauge theory on $S^5$, 
including background supergravity construction for the vector multiplets.

\section{Coulomb branch indices for the self-dual strings}

In this section, we study the spectrum of self-dual strings in the Coulomb phase of
the 6d SCFTs. On one hand, this will be interesting data of the theory by itself.
On the other hand, these Coulomb phase observables play important roles in understanding 
supersymmetric partition functions at the conformal point, such as the 6d superconformal index
\cite{Bhattacharya:2008zy}.

In the `Coulomb phase,' scalars in the 6d tensor multiplets assume nonzero expectation 
values $v^I$. In such a phase, there appear tensionful self-dual strings whose tension is
proportional to the Coulomb VEV. Let us first explain the SUSY preserved by these strings,
when they are extended along a straight line.
The 6d theory in the Coulomb phase preserves $\mathcal{N}=(1,0)$ or $(2,0)$ 
Poincare supersymmetry. We only use the $(1,0)$ part of the SUSY to define our BPS self-dual
strings. For our purpose, we write the $8$ supercharges
as $Q^A_{\alpha}$, $Q^A_{\dot\alpha}$. $A=1,2$ is
the doublet index for the $SU(2)_R$ R-symmetry. $\alpha=1,2$, $\dot\alpha=1,2$ are the
doublet indices of the $SU(2)_l\times SU(2)_r=SO(4)$ spatial rotation on the 6d field 
theory direction $\mathbb{R}^4$, transverse to the string. These supercharges are subject 
to the reality condition
\begin{equation}\label{KLreality}
  (Q^A_\alpha)^\dag=\epsilon_{AB}\epsilon^{\alpha\beta}Q^B_{\beta}\ ,\ \
  (Q^A_{\dot\alpha})^\dag=\epsilon_{AB}\epsilon^{\dot\alpha\dot\beta}Q^B_{\dot\beta}\ .
\end{equation}
The supersymmetry algebra contains the following
anti-commutatiaon relations: 
\begin{eqnarray}\label{KLalgebra}
  &&\{Q^A_\alpha,Q^B_\beta\}=\epsilon^{AB}\epsilon_{\alpha\beta}\left(H+P+Rv^In_I\right),\
  \{Q^A_{\dot\alpha},Q^B_{\dot\beta}\}=\epsilon^{AB}\epsilon_{\dot\alpha\dot\beta}
  \left(H-P-Rv^In_I\right),\nonumber\\
  &&\{Q^A_{\alpha},Q^B_{\dot\beta}\}=\epsilon^{AB}(\sigma^m)_{\alpha\dot\beta}P_m\ ,
\end{eqnarray}
where $H$ is energy, $P$ is momentum along the string, and $P_m$ is the momenta along $\mathbb{R}^4$.
Here we have compactified one direction of the 6d theory on $S^1$ with radius $R$, and wrapped
the strings on that circle. We shall study the self-dual strings whose 5d masses saturate
the BPS bound $H\geq P+Rv^In_I$, so preserve $4$ supercharges $Q^A_{\dot\alpha}$.
The other half-BPS states preserving $Q^A_{\alpha}$ would have similar spectrum.

\subsection{Elliptic Genus Method}

In particular, we shall be interested in the Witten index which
counts the BPS degeneracies of these strings wrapping the circle. Namely, the 6d CFT is
put on $\mathbb{R}^{4,1}\times S^1$, and there are $r$ real scalar
VEVs $v^I$ ($I=1,\cdots,r$). The index is defined by
\begin{equation}
  Z_{\{n_I\}}(\tau,\epsilon_{1,2},m)={\rm Tr}\left[(-1)^Fq^{\frac{H'+P}{2}}
  e^{-\epsilon_1(J_1+J_R)-\epsilon_2(J_2+J_R)}e^{-m\cdot F}\right]\ ,
\end{equation}
where $H'$ is the energy over the string rest mass $Rv^I n_I$, $q\equiv e^{2\pi i\tau}$, 
$J_1\equiv J_l+J_r$, $J_2\equiv J_r-J_l$, and $F$ collectively denotes all the other 
conserved global charges which commute with the supercharges.
The charges appearing inside the trace is chosen so that they commute with the two
supercharges $Q^1_{\dot{1}}$, $Q^2_{\dot{2}}$, among $Q^A_{\dot\alpha}$.
From the algebra (\ref{KLalgebra}), the most
general states preserving these two supercharges will be the $\frac{1}{2}$-BPS states
preserving all $Q^A_{\dot\alpha}$. So with this index we are counting the states in the
$\frac{1}{2}$-BPS multiplet, with a further refinement given by $J_r+J_R$ (which does
not commute with all four $Q^A_{\dot\alpha}$). We also define the partition function
$Z(v^I,\tau,\epsilon_{1,2},m)$ by summing over
the winding numbers of the self-dual strings,
\begin{equation}\label{KL2d-expansion}
  Z(v^I,\tau,\epsilon_{1,2},m)=\sum_{n_1,\cdots,n_r=0}^\infty e^{-v^In_I}
  Z_{n_I}(\tau,\epsilon_{1,2},m)\ , 
\end{equation}
where $Z_{n_I=0}\equiv 1$. Here, we introduce the (dimensionless) chemical 
potentials $v^I$ conjugate to the winding numbers $n_I$. These are just scaled version of   
the scalar VEVs $v^I$ that we used above but should not be confused with them.

$Z(v^I,\tau,\epsilon_{1,2},m)$ is computed in various ways. Currently, in most
nontrivial theories, it is only computable in series expansions. One series expansion
takes the form of (\ref{KL2d-expansion}), and the coefficients $Z_{n_I}(\tau,\epsilon_{1,2},m)$
are computed from the elliptic genera of suitable 2 dimensional supersymmetric quantum
field theories living on the worldsheets of these strings
\cite{Haghighat:2013gba,Haghighat:2013tka,Kim:2014dza,Haghighat:2014vxa}.
A different kind of series expansion can be made in $q=e^{2\pi i\tau}$, when $q\ll 1$:
\begin{equation}\label{KL1d-expansion}
  Z(v^I,\tau,\epsilon_{1,2},m)=\sum_{k=0}^\infty q^kZ_k(v^I,\epsilon_{1,2},m)\ .
\end{equation}
The momentum charge $k$ on $S^1$ is given a
weight $q^{k}$. These Kaluza-Klein momentum states are
regarded as massive particles in 5d. $Z_k(v^I,\epsilon_{1,2},m)$ can be computed from
the quantum mechanics of the `instanton solitons' of 5 dimensional gauge theory, if one
has a 5d weakly coupled SYM description at small radius. In this section, we shall mostly
focus on the latter quantum mechanical index. The usefulness of these two
approaches will be commented later.

We first explain the general ideas of computing the two types of coefficients
$Z_{n_I}(q,\epsilon_{1,2},m)$ and $Z_k(v^I,\epsilon_{1,2},m)$, before studying
an example. Both computations essentially rely on the string
theory completion of the 6d SCFT, and suitable decoupling limits when the contribution
of some charges to the BPS mass become large.

Let us first explain the strategy of computating $Z_{n_I}(q,\epsilon_{1,2},m)$.
Firstly, as the 6d SCFT lacks intrinsic definition, we rely on its string theory or
M-theory engineering. In all such constructions, one engineers suitable string/M-theory
backgrounds, and takes suitable low energy decoupling limits in
which the 6 dimensional states decouple from the bulk states (e.g. 10/11 dimensional gravity, stringy states, so on). After
this limit, certain 6 dimensional decoupled sector of 6d SCFT exists.
Furthermore, we are interested in the 1+1 dimensional strings in the Coulomb phase,
with nonzero VEV for the 6d scalar $v$ whose mass dimension is $2$. The tension of
the self-dual strings is proportional to $v$. At energy scale much below $v^{\frac{1}{2}}$,
the 6d system will again exhibit a decoupling, between the 2d QFT on the strings
and the rest of the 6d system. $Z_{n_I}(q,\epsilon_{1,2},m)$ is computed by studying
the last 2d QFT living on the strings' worldsheet. We generally expect
the 2d QFT to be an interacting conformal field theory. The computation of
the observables is generally very difficult with strongly interacting QFT. Here, the
crucial step is to engineer a 2d gauge theory which is weakly coupled in UV,
and flows to the desired interacting CFT in the IR. The construction of the UV gauge
theory will often be easy with brane construction engineering of the 6d SCFT and the
associated self-dual strings. Such UV gauge theories are constructed for the self-dual 
strings of a few interesting 6d CFTs, such as `M-strings' \cite{Haghighat:2013tka}, 
`E-strings' \cite{Kim:2014dza}, and some others \cite{Haghighat:2014vxa}.
The UV gauge theories for many interesting
self-dual strings are still unknown at the moment and are under    active studies.
With a weakly-coupled UV gauge theory which flows to the desired CFT, the
elliptic genus can in principle be easily computed from the UV theory, as the
elliptic genus is independent of the continuous coupling parameters of the theory.
In fact the general elliptic genus formula for 2d SUSY Yang-Mills theories was 
recently derived in \cite{Benini:2013nda,Benini:2013xpa}.

$Z_{k}(v^I,\epsilon_{1,2},m)$ can also be computed in a similar manner, 
for some classes of self-dual strings. This approach is applicable to the cases in which
the circle compactification of the 6d theory yields weakly coupled 5d Yang-Mills theories
at low energy. Then, the momentum $k$ is given by the topological charge
\begin{equation}
  k=\frac{1}{8\pi^2}\int_{\mathbb{R}^4}{\rm tr}(F\wedge F)\in\mathbb{Z}
\end{equation}
carried by the Yang-Mills instanton solitons of the 5d gauge theory. The dynamics
of these solitons are often described by a quantum mechanical gauge theory.
$Z_k( v^I,\epsilon_{1,2},m)$ is essentially computed by the quantum mechanical
index for the $k$ instantons. More precisely, one finds
\begin{equation}
  Z_k(v^I,\epsilon_{1,2},m)=Z_{\rm pert}(v^I,\epsilon_{1,2},m)
  Z_{k,{\rm inst}}(v^I,\epsilon_{1,2},m)\ ,
\end{equation}
where $Z_{\rm pert}$ is computed from the perturbative degrees of freedom in 5d SYM,
and $Z_{k,{\rm  inst}}$ is given by the instanton quantum mechanics. The last instanton
partition function has been first computed in \cite{Nekrasov:2002qd,Nekrasov:2003rj},
and has been intensively studied since then for various reasons. Although we used the
notion of 5d SYM to explain the strategy, we can often get the quantum mechanical
gauge theory description from the full string theory set up by taking a suitable
decoupling limit, bypassing the UV incomplete 5d SYM description at all. For instance,
for the $(2,0)$ theory compactified on circle, one just obtains the quantum mechanics
from the D0-D4 system by taking a low energy decoupling limit, without relying on
5d SYM description at all.

The two quantities $Z_{n_I}(q,\epsilon_{1,2},m)$ and $Z_{k}(v^I,\epsilon_{1,2},m)$
are supersymmetric indices of the 2d and 1d gauge theories on $T^2$ and $S^1$, respectively.
Although both types of indices have been extensively studied in the literature from long
time ago, their general structures for gauge theories have been fully clarified only recently. See \cite{Gadde:2013ftv,Benini:2013nda,Benini:2013xpa} for the developments in the 2d elliptic genus, and
\cite{Hwang:2014uwa,Cordova:2014oxa,Hori:2014tda} for the 1d Witten index.

Before proceeding with concrete examples, we also comment that the quantity
$Z(v^I,\tau,\epsilon_{1,2},m)$ can often be computed from topological
string amplitudes on suitable Calabi-Yau 3-folds. This happens when
the 6d SCFTs are engineered from F-theory on singular
elliptic Calabi-Yau 3-folds \cite{Morrison:1996na,Morrison:1996pp,Witten:1996qb}.
Changing the moduli of CY$_3$ in a
way that specific 2-cycles shrink to zero volume, one obtains a 6 dimensional singularity
which supports decoupled degrees of freedom at low energy, defining 6d SCFTs. One important
ingredient of these theories is D3-branes wrapping these collapsing 2-cycles, which yield
self-dual strings that become tensionless in the singular limit. Therefore, the volume
moduli of these 2-cycles are the Coulomb branch VEVs $v^I$ in the 6d tensor supermultiplets.

So in this setting, we consider the F-theory on
$\mathbb{R}^{4,1}\times S^1\times {\rm CY}_3$ in the Coulomb phase. 
We wrap D3-branes along $S^1$ times the 2-cycles in CY$_3$.
This system can be T-dualized on $S^1$ to the dual circle $\tilde{S}^1$ of the type IIA
theory. The D3-branes map to D2-branes transverse to $\tilde{S}^1$. Consider
the regime with large $S^1$, or equivalently small $\tilde{S}^1$, and make an M-theory
uplift on an extra circle $S^1_M$. Then $\tilde{S}^1$ and $S^1_M$ combine to a torus and fiber
the 4d base of the original CY$_3$ we started from, meaning that we get M-theory on the 
same CY$_3$. The self-dual string winding numbers over the 2-cycles maps to
the M2-brane winding numbers on the same cycles. The momentum on $S^1$ maps to M2-brane winding
number on the $T^2$ fiber. So the counting of the self-dual string states
maps to counting the wrapped M2-branes on CY$_3$ in M-theory. The last BPS spectrm is
computed by the topological string partition function on CY$_3$
\cite{Gopakumar:1998ii,Gopakumar:1998jq}.
In particular, consider an expansion of $Z_{\mathbb{R}^4\times T^2}$
in the rotation paramters $\epsilon_1,\epsilon_2$ given by
\begin{equation}\label{KLgenus}
  Z_{\mathbb{R}^4\times T^2}(v^I,q,\epsilon_{1,2},m)=
  \exp\left[\sum_{n\geq 0,g\geq 0}
  (\epsilon_1+\epsilon_2)^n(\epsilon_1\epsilon_2)^{g-1}F^{(n,g)}(v^I,q,m)\right]\ .
\end{equation}
The coefficients of the expansion $F^{(n,g)}(v^I,q,m)=
\sum_{n_I,k,f}e^{-v^In_I}q^ke^{-m\cdot f}F^{(n,g)}_{n_I,k,f}$ are
computed by the topological string amplitudes on CY$_3$. The series in
(\ref{KLgenus}) is the genus expansion of refined topological string. So from this viewpoint, 
the elliptic genus we study in this section is the all genus sum of the topological string 
amplitudes. A few low genus expansions are known for many interesting 6d self-dual strings. 
This provides an alternative method of computing some data of the full elliptic
genus when neither 2d nor 1d gauge theories are known. For instance, see \cite{Haghighat:2014vxa} 
for the results 6d strings engineered by F-theory on Hirzebruch surfaces, where many such 
strings do not have known gauge theory descriptions  yet.

\subsection{Instanton Partition Method}

With the above comments in mind, we shall now explain the studies of the Coulomb branch
indices from 1d gauge theories. We shall specifically explain the 6d $(2,0)$ SCFT of
$A_{N-1}$ type, to be concrete. Although this quantity has been studied in the context 
of `instanton counting' of 5d SYM \cite{Nekrasov:2002qd},
one does not have to rely on 5d SYM description at all, as everything can be directly
understood from the full string theory setting. For applications of the similar techniques
to the Coulomb branch CFTs with $(1,0)$ SUSY, see \cite{Hwang:2014uwa,Kim:2014dza}.

The maximal superconformal field theory in 6d of $A_{N-1}$ type is engineered by
taking $N$ M5-branes on top of another, in the flat M-theory background. In the low energy
limit, the system contains a 6d SCFT on M5-branes' worldvolume which is decoupled from
the bulk. In the Coulomb branch, we take $N$ M5-branes separated along one of the five
transverse directions of $\mathbb{R}^5$. The self-dual strings are suspended between
separated M5-branes along this direction, and also wrap $\mathbb{R}^{1,1}\subset \mathbb{R}^{5,1}$
of the 5-brane worldvolume.

We are interested in the index of the circle compactified self-dual strings.
The index is invariant under the change of continuous parameters of the theory,
and also of the background parameters as long as they do not appear in the supercharges
that are associated with the definition of the Witten index. So we can take the circle
radius to
be very small, and use the type IIA string theory description for the computation. 
Let us denote by $v^I$ (with $I=1,\cdots,N$) the $N$ scalar VEVs, or positions of 
$N$ M5-branes along a line in $\mathbb{R}^5$. These are related to the 5d VEVs  by a multiplication of $R$. Let $n_I$ denote the number of self-dual 
strings ending on a given M5-branes, with orientations taken into account.
If the strings have $k$ units of Kaluza-Klein momentum, one obtains in the small $R$ limit 
a system of $k$ D0-branes bound to fundamental strings with charges $n_I$ stretched 
between the $N$ D4-branes. In particular, the energy of
the compactified self-dual strings is bounded as
\begin{equation}
  E\geq\frac{k}{R}+v^I n_I\ .
\end{equation}
In the regime with very small $R$, where we plan to compute the index, we can use
the effective description with fixed $k$, as the particles with large rest mass $\sim R^{-1}$
become non-relativistic. So the quantum mechanics of $k$ D0-branes bound to $N$ D4-branes
would capture the exact index $Z_{k,{\rm inst}}(v^I,\epsilon_{1,2},m)$. The quantum
numbers $n_I$ will be realized as $SU(N)$ Noether charges of this mechanical system.
This is simply the decoupling limit of the $k$ D0-branes bound to D4-branes, and could 
also be regarded as the discrete lightcone quantization (DLCQ) of M5-branes \cite{Aharony:1997th}.

The quantum mechanics of $k$ D0-branes on $N$ D4-branes (in the Coulomb phase) preserves
$8$ SUSY, since the D0-D4 system preserves $\frac{1}{4}$ of the type IIA SUSY. The system
has $SO(4)_1\sim SU(2)_{1L}\times SU(2)_{1R}$ rotation symmetry on D4 worldvolume
transverse to D0, and $SO(5)$ rotation transverse to the D4's. When D4's are displaced 
along one of the five directions of $\mathbb{R}^5$, with VEV $v={\rm diag}(v_1,\cdots,v_N)$, 
$SO(5)$ is broken to $SO(4)_2\sim SU(2)_{2L}\times SU(2)_{2R}$. We denote by 
$\alpha,\dot\alpha,a,\dot{a}$ the doublet indices of the four $SU(2)$'s,
respectively, in the order presented above. The $8$ supercharges can be written by
$Q^a_{\dot\alpha}$, $Q^{\dot{a}}_{\dot\alpha}$ with reality
conditions similar to (\ref{KLreality}). The degrees
of freedom are:
\begin{eqnarray}
  \textrm{D0-D0 strings}&:&U(k)\textrm{ adjoint }A_0\ ,\ \ 
  (\varphi^{1,2,3,4}\sim\varphi_{a\dot{a}},\varphi^5)\ ,\ \
  \lambda^{a}_{\dot\alpha}\ ,\ \ \lambda^{\dot{a}}_{\dot\alpha}\nonumber\\
  &&U(k)\textrm{ adjoint }a_m\sim a_{\alpha\dot\alpha}\ ,\ \
  \lambda_{\alpha}^a\ ,\ \ \lambda^{\dot{a}}_\alpha\nonumber\\
  \textrm{D0-D4 strings}&:&U(k)\times U(N)\textrm{ bi-fundamental }
  q_{\dot\alpha}\ ,\ \ \psi^a\ ,\ \ \psi^{\dot{a}}
\end{eqnarray}
with $m=1,\cdots,4$.
The D4-D4 strings move along $\mathbb{R}^4$ transverse to the D0's, and decouple
at low energy. (These will be perturbative 5d SYM degrees.)
This system can be formally obtained by a dimensional reduction of a 2 dimensional
$\mathcal{N}=(4,4)$ SUSY gauge theory, in which $Q^a_{\dot\alpha}$ and
$Q^{\dot{a}}_{\dot\alpha}$ respectively define $4$ left-moving and right-moving
supercharges. The first line of
the above field content is called the vector multiplet. The second and third line
separately form a hypermultiplet. The action of this system is very standard, and
could be found e.g. in \cite{Kim:2011mv}, whose notations we followed here.

The index $Z_{k,{\rm inst}}(v^I,\epsilon_{1,2},m)$ is defined in this quantum
mechanics by
\begin{equation}
  Z_{k,{\rm inst}}(v^I,\epsilon_{1,2},m)={\rm Tr}\left[
  (-1)^Fe^{-\beta\{Q,Q^\dag\}}
  e^{-v\cdot n}e^{-2\epsilon_+(J_{1R}+J_{2R})}e^{-2\epsilon_-J_{1L}}
  e^{-mJ_{2L}}\right]\ ,
\end{equation}
where $n=(n_1,\cdots,n_N)$ denotes the $U(1)^N\subset U(N)$ charge,
$\epsilon_\pm\equiv\frac{\epsilon_1\pm\epsilon_2}{2}$, and $J_{1L},J_{1R}$,
$J_{2L},J_{2R}$ are the Cartans of $SU(2)_{1L},SU(2)_{1R}$, $SU(2)_{2L},SU(2)_{2R}$,
respectively. $\beta$ is the usual regulator parameter which does not appear in the index.
The trace is over the Hilbert space of the quantum mechanics. Note that the measure in
the trace commutes with two supercharges $Q=Q^{\dot{1}}_{\dot{1}}$,
$Q^\dag=-Q^{\dot{2}}_{\dot{2}}$
among $Q^{\dot{a}}_{\dot\alpha}$, as we explained at the beginning of this section.
This index was computed by Nekrasov \cite{Nekrasov:2002qd} in 2002. We shall briefly
review it with adding more recent clarifications on the computational step, for which
Nekrasov wrote down a prescription for computation. These clarification of the
prescriptions is somewhat crucial to compute the $Z_{k,{\rm inst}}$ indices for more
general 6d $(1,0)$ SCFTs \cite{Hwang:2014uwa,Kim:2014dza}.

The 1d gauge theory for D0-D4 system that we explained above is strongly coupled
at low energy, since the quantum mechanical gauge coupling has dimension $[g_{QM}^2]=M^3$. We
can however compute the index in the $g_{QM}\rightarrow 0$ limit, as the Witten index
is generically expected to be insensitive to the changes of continuous parameters of
the theory. This is what Nekrasov has done in \cite{Nekrasov:2002qd}, and also in more
recent studies of \cite{Hwang:2014uwa,Cordova:2014oxa,Hori:2014tda}.
The computation
of the index is done by going to the path integral representation of the index
with Euclidean quantum mechanics, put on a circle with circumference $\beta$, and
computing it in the $g_{QM}\rightarrow 0$ limit. The computation
consists of (1) identifying the zero modes of the quantum mechanical path
integral on $S^1$, in the limit $g_{QM}\rightarrow 0$ (carefully defined in
\cite{Benini:2013nda,Benini:2013xpa,Hwang:2014uwa,Hori:2014tda}); (2) Gaussian path
integral over the non-zero modes; (3) finally making an
exact integration over the zero modes.

We briefly explain the results of these three steps, within our example for simplicity. 
Firstly, the zero modes in the
$g_{QM}\rightarrow 0$ limit consist of the constant modes of $\varphi^5$
and $A_\tau$ which commute with each other. Here, $A_\tau$ is the Wick-rotated variable $A_\tau=-iA_0$ in the Euclidean quantum mechanics on $S^1$. More precisely,
$U\equiv e^{i\beta A_\tau}$ defines a holonomy of the gauge group $U(k)$ along the
circle. For the $U(k)$ gauge group, one can take $\phi\equiv\varphi^5+iA_\tau$ to be
\begin{equation}\label{KLQM-0-mode}
  \phi={\rm diag}(\phi_1,\cdots,\phi_k)
\end{equation}
using $U(k)$ rotation, locally labeled by $k$ complex parameters. Each parameter satisfies
$\phi_i\sim\phi_i+2\pi i$, so lives on a cylinder. These variables are subject
to further identification given by permuting the $k$ variables. This is the
permutation subgroup of $U(k)$ which acts within (\ref{KLQM-0-mode}). For gauge groups 
other than $G=U(k)$, especially for disconnected groups, the zero mode structure could 
be more complicated. See \cite{Hwang:2014uwa} for examples. There are also some
fermionic zero modes in the strict $g_{QM}=0$ limit, which we shall not explain here, 
but plays important roles in the final step (3) above.

Secondly, in the above
background, the 1-loop determinants over non-zero modes yield the following factor:
\begin{eqnarray}\label{KL1-loop}
  Z_{\textrm{1-loop}}(\phi,\epsilon_{1,2},m)&=&  \frac{\prod_{I\neq J}  2\sinh\frac{\phi_{IJ}}{2}\cdot\prod_{I,J=1}^k2\sinh\frac{\phi_{IJ}+2\epsilon_+}{2}}
  {\prod_{I,J=1}^k   2\sinh\frac{\phi_{IJ}+\epsilon_1}{2}\cdot 2\sinh\frac{\phi_{IJ}+\epsilon_2}{2}    }
  \cdot
  \prod_{I,J=1}^k\frac{2\sinh\frac{\phi_{IJ}\pm m-\epsilon_-}{2}}
  {2\sinh\frac{\phi_{IJ}\pm m-\epsilon_+}{2}}\nonumber\\
  &&\cdot\prod_{I=1}^k\prod_{i=1}^N\frac{2\sinh\frac{m\pm(\phi_I-v_i)}{2}}
  {2\sinh\frac{\epsilon_+\pm(\phi_I-v_i)}{2}} \ , 
\end{eqnarray}
where $\phi_{IJ}\equiv\phi_I-\phi_J$, and the $\sinh$ expressions with $\pm$ in the
arguments mean multiplying the $\sinh$ factors with all possible signs. The factor on
the second line comes from the integral over the fundamental hypermultiplet
$q_{\dot\alpha},\psi^a,\psi^{\dot{a}}$, and the second factor on the right hand side
of the first line comes from the adjoint hypermultiplet
$a_m,\lambda^a_\alpha,\lambda^{\dot{a}}_\alpha$. Finally, the first factor on the
first line comes from the vector multiplet nonzero modes
$A_\tau,\varphi^I,\lambda^a_{\dot\alpha},\lambda^{\dot{a}}_{\dot\alpha}$.

The final task is to integrate over the $k$ complex, or $2k$ real, variables $\phi_I$.
Naively, it appears that one has to do a $2k$ dimensional integral on copies of cylincers,
with a meromorphic measure given by (\ref{KL1-loop}). This naive prescription will not work
because the measure will diverge at various poles, implying that the integration over non-zero modes becomes subtle near the poles even in the $g_{QM}\rightarrow 0$ limit. In
\cite{Nekrasov:2002qd}, Nekrasov gave a $k$ dimensional contour integral
prescription, rather than a $2k$ dimensional real integral, with the measure (\ref{KL1-loop}).
The result is the sum over residues for a subset of poles in the integrand (\ref{KL1-loop}).
The relevant poles are labeled by all possible $N$-tuple of Young diagrams
$Y=(Y_1,Y_2,\cdots,Y_N)$ with $k$ total
number of boxes. These are sometimes called $N$-colored Young diagrams with $k$ boxes. 
The summation of residues from these poles is given by 
\cite{Flume:2002az,Bruzzo:2002xf,Kim:2011mv}
\begin{equation}\label{KLmaximal-instanton-index}
  Z_{k,{\rm inst}}(v,\epsilon_{1,2},m)=\sum_{|Y|=k}
  \prod_{i,j=1}^N\prod_{s\in Y_i}\frac{\sinh\frac{E_{ij}+m-\epsilon_+}{2}
  \sinh\frac{E_{ij}-m-\epsilon_+}{2}}
  {\sinh\frac{E_{ij}}{2}\sinh\frac{E_{ij}-2\epsilon_+}{2}}
\end{equation}
with
\begin{equation}
  E_{ij}=v_i-v_j-\epsilon_1h_i(s)+\epsilon_2(v_j(s)+1)\ .
\end{equation}
Here, $s$ labels the boxes in the $i$'th Young diagram $Y_i$. $h_i(s)$ is the distance
from the box $s$ to the edge on the right side of $Y_i$ that one reaches by moving
horizontally to the right. $v_j(s)$ is the distance from $s$ to the edge on the bottom side
of $Y_j$ that one reaches by moving down (and $v_j(s)$ may be negative if one has to move
up to the bottom of $Y_j$). See \cite{Flume:2002az,Bruzzo:2002xf,Kim:2011mv} for more
detailed explanations on notation. This result can be obtained from the following rule
for the contour. First of all, the contours will be a closed curve on the
$z_I=e^{\phi_I}$ plane. The rules of the contour choices, or equivalently the residues
to be kept by the contour integral, are as follows: (1) exclude all the poles in
(\ref{KL1-loop}) coming from the $\sinh$ factors whose arguments include $m$ (from 5d SYM,
this amounts to ignoring all the poles coming from 5d adjoint hypermultiplet); (2)
exclude all poles at $z_I=0$ or $\infty$; (3) as for the remaining poles, take 
$\epsilon_+\gg 1$, and include all poles within the unit circles $|z_I|<1$.  

Although well known and used, this prescription was given a satisfactory derivation only rather recently \cite{Hwang:2014uwa},
using and generalizing the methods of \cite{Benini:2013nda,Benini:2013xpa}. The strategy 
of \cite{Benini:2013nda,Benini:2013xpa} 
is to carefully re-do the supersymmetric path integral computation when $\phi$ is near
its pole location, and also carefully considering the lift of some gaugino zero modes
\cite{Benini:2013nda}. After some analysis, the final contour integral reduces to a set
of residue sum, which is called the Jeffrey-Kirwan residue \cite{JK}. The Jeffrey-Kirwan
residue rules are slightly different from the above (1), (2), (3) in general, but it was
shown for the above $U(k)$ theory that the two rules yield the same result \cite{Hwang:2014uwa}.

The result (\ref{KLmaximal-instanton-index}) is useful to understand various aspects of
the $(2,0)$ theory in Coulomb phase, and its self-dual strings comapactified on a circle
\cite{Kim:2011mv}. It is also useful to understand the conformal   phase (with zero Coulomb
VEV) of the theory. An early finding of this sort was that
(\ref{KLmaximal-instanton-index}) could be used to study the index of the DLCQ $(2,0)$ theory,
which is the 6d CFT compactified on a light-like circle. Namely, one takes
(\ref{KLmaximal-instanton-index}) and suitably integrates over the Coulomb VEV $v$ with
Haar measure inserted, to extract out the gauge invariant spectrum \cite{Kim:2011mv}.
More recently, and this will be reviewed in our section 3, (\ref{KLmaximal-instanton-index})
was used as the building block of more sophisticated CFT observable, the superconformal
index on $S^5\times S^1$. Again several factors of the form (\ref{KLmaximal-instanton-index})
are multiplied (with other factors that we shall call the `classical measure,' see section 3), and we suitably integrate over the Coulomb VEV parameter $v$.

Here, we find one virtue of the index (\ref{KL1d-expansion}) obtained by 1d gauge theory,
over (\ref{KL2d-expansion}) which is obtained by 2d gauge theory. Namely, in many recent
applications, $q=e^{2\pi i\tau}$ is kept as a fixed fugacity, while the Coulomb VEV is
introduced temporarily and should be integrated over to obtain CFT observables. The
computations explained in this subsection keeps the $v$ dependence exact, at a given
order $q^k$ in $q$. So in this sense, knowing the coefficients of  (\ref{KL1d-expansion}) 
exactly could be more useful, rather than knowing those of (\ref{KL2d-expansion}).

On the other hand, the elliptic genus (\ref{KL2d-expansion}) has the virtue of making
the modular property under the $SL(2,\mathbb{Z})$ transformation clear, with the
modular parameter $\tau$. So when one has to make a strong coupling re-expansion of 
the partition function, as explained in the introduction and section 3.2, this could 
potentially be very useful. Also, the elliptic genus (\ref{KL2d-expansion}) can often be
computed when the circle reduction of 6d CFT does not flow to weakly coupled SYM,
so that the 1d approach of this section becomes difficult to apply
\cite{Kim:2014dza,Haghighat:2014vxa}. However, (\ref{KL2d-expansion}) takes the form of 
the Coulomb VEV expansion when $v$ acquires large expectation values (compared to other 
parameters such as $\tau,m,\epsilon_{1,2}$). So apparently it is unclear how to integrate 
over them in the curved space partition functions.

\section{Superconformal indices of the 6d $(2,0)$ theories}

In this section, we explain the current status of our understanding on the superconformal
index of the 6d $(2,0)$ theories. Possible extensions to the 6d $(1,0)$ theories have not
 yet  been developed in detail, on which we shall just make general statements and brief comments.

6d $(1,0)$ SCFT has $OSp(8^\ast|2)$ superconformal symmetry, as well as
possible global symmetries whose charges we collectively call $F$. The bosonic part
of the superconformal symmetry is $SO(6,2)\times SU(2)_R$. We are interested in the
radially quantized CFT, living on $S^5\times \mathbb{R}$. Then the maximal commuting set
of charges of the bosonic subgroup are taken to be $R\in SU(2)_R$ in the R-symmetry,
$j_1,j_2,j_3\in SO(6)$ which are rotations on $S^5$, and $E$ for the translational
symmetry along the time direction $\mathbb{R}$. Often, we make $E$ dimensionless by
multiplying the radius $r$ of $S^5$. We normalize $R,j_i$ to have $\pm\frac{1}{2}$
eigenvalues for spinors. The superconformal index of a general 6 dimensional SCFT
is defined by \cite{Bhattacharya:2008zy}
\begin{equation}\label{KLSC-index}
  Z_{S^5\times S^1}(\beta,m,a_i)\equiv{\rm Tr}
  \left[(-1)^Fe^{-\beta(E-R)}e^{-\beta a_i j_i}e^{-\beta m\cdot F}\right]\ ,
\end{equation}
where $a_1+a_2+a_3=0$, and ${\rm Tr}$ is the trace over the Hilbert space of the CFT on
$S^5\times\mathbb{R}$. Note that the 6d $(1,0)$ SCFT has
$8$ Poincare supercharges $Q^A_{s_1s_2s_3}$, with $A=\pm\frac{1}{2}$ for the R-symmetry, and
$(s_1,s_2,s_3)=(\pm\frac{1}{2},\pm\frac{1}{2},\pm\frac{1}{2})$ for the $SO(6)$ symmetry, where
the last three $\pm$ signs are constrained by $\pm\pm\pm =-$. These supercharges have energy
($\sim$ scale dimension) $E=\frac{1}{2}$. The $8$ conformal supercharges
are given by $S^A_{s_1s_2s_3}$, where this time the three signs for $s_i=\pm\frac{1}{2}$ are
constrained by $\pm\pm\pm=+$. They have scale dimension $E=-\frac{1}{2}$. Among these $16$
supercharges, the measure of (\ref{KLSC-index}) commutes with $Q=Q^{+}_{---}$,
$S=S^-_{+++}$. So the index counts BPS states (with minus sign for fermions) which are
annihilated by at least these two supercharges. Equivalently, by the operator-state map,
the index counts BPS local operators of the CFT on $\mathbb{R}^6$. The energies (dimensions) 
of the BPS states (operators) are given by $E=4R+j_1+j_2+j_3$, from the vanishing of $\{Q,S\}$ 
acting on these BPS states.

Specifying to the 6d $(2,0)$ SCFTs, the superalgebra is $OSp(8^\ast|4)$, and there are no 
extra flavor symmetries. The supercharges are now given
by $Q^{R_1R_2}_{s_1s_2s_3}$, $S^{R_1R_2}_{s_1s_2s_3}$, where $R_1=\pm\frac{1}{2}$,
$R_2=\pm\frac{1}{2}$ are the two $SO(5)\sim Sp(4)$ spinor charges. $s_1,s_2,s_3$ are given
and constrained in the same way as the previous paragraph. We can pick
$Q=Q^{++}_{---}$ and $S=S^{--}_{+++}$ and define the index
\begin{equation}\label{KLSC-index-(2,0)}
  Z_{S^5\times S^1}(\beta,m,a_i)\equiv{\rm Tr}\left[(-1)^F
  e^{-\beta(E-\frac{R_1+R_2}{2})}e^{-\beta a_i j_i}e^{\beta m\frac{R_1-R_2}{2}}\right]\ .
\end{equation}
The BPS states counted by this index satisfy $E=2(R_1+R_2)+j_1+j_2+j_3$.
(\ref{KLSC-index-(2,0)}) can be regarded as a specialization of (\ref{KLSC-index}) by
regarding $R=\frac{R_1+R_2}{2}$ as the $(1,0)$ R-charge, and $\frac{R_1-R_2}{2}$
as a flavor symmetry of the $(1,0)$ superconformal subalgebra.

Even without a microscopic formulation of the 6d SCFTs, we have fairly well-motivated
expressions for these partition functions (\ref{KLSC-index}), (\ref{KLSC-index-(2,0)}).
We shall write down
two such expressions, one in section 3.1 and another in section 3.3. Both of them are
inspired by 5 dimensional super-Yang-Mills theories, obtained by circle
reductions of 6d SCFTs on $S^5\times S^1$ down to 5d.

\subsection{The partition function on $S^5$}

The first expression for the partition function $Z_{S^5\times S^1}$,
is given as follows. It uses the Coulomb branch partition function
$Z_{\mathbb{R}^4\times T^2}(\tau,v,\epsilon_1,\epsilon_2,m_0)$ that we explained in
section 2, and is given by
\begin{eqnarray}\label{KLSC-index-1}
  \hspace*{-1 cm}Z_{S^5\times S^1}(\beta,m,a_i)&=&\frac{e^{-S_{\rm bkgd}}}{|W(G_r)|}
  \int_{-\infty}^\infty\left[\prod_{I=1}^rd\phi_I\right]e^{-S_0(\phi,\beta,a_i)}
  Z_{\mathbb{R}^4\times T^2}\left(\frac{2\pi i}{\beta\omega_1},\frac{\phi}{\omega_1},
  \frac{2\pi i\omega_{21}}{\omega_1},\frac{2\pi i\omega_{31}}{\omega_1},
  2\pi i\left(\frac{m}{\omega_1}+\frac{3}{2}\right)\right)\nonumber\\
  &&\hspace{-3cm}\cdot Z_{\mathbb{R}^4\times T^2}\left(\frac{2\pi i}{\beta\omega_2},
  \frac{\phi}{\omega_2},\frac{2\pi i\omega_{32}}{\omega_2},
  \frac{2\pi i\omega_{12}}{\omega_2},
  2\pi i\left(\frac{m}{\omega_2}+\frac{3}{2}\right)\right)
  Z_{\mathbb{R}^4\times T^2}\left(\frac{2\pi i}{\beta\omega_3},\frac{\phi}{\omega_3},
  \frac{2\pi i\omega_{13}}{\omega_3},\frac{2\pi i\omega_{23}}{\omega_3},
  2\pi i\left(\frac{m}{\omega_3}+\frac{3}{2}\right)\right), \nonumber\\
  \hspace*{-1.5cm}S_0&=&\frac{2\pi^2{\rm tr}(\phi^2)}{\beta\omega_1\omega_2\omega_3}\ ,
\end{eqnarray}
where $\omega_i\equiv 1+a_i$, $\omega_{ij}\equiv\omega_i-\omega_j=a_i-a_j$.
($\frac{2\pi i\omega_{ij}}{\omega_j}$ appearing in the arguments may be replaced by 
$\frac{2\pi i\omega_i}{\omega_j}$, as was more commonly used in \cite{Lockhart:2012vp,Qiu:2013aga}, 
using the $2\pi i$ period shifts of the arguments.)
Here, $G_r$ is the gauge group of the low energy 5d SYM that one obtains by reducing the
6d SCFT, and $W(G_r)$ is the Weyl group of $G_r$. More abstractly, in the 6d CFT, 
$W(G_r)$ acquires meaning as the Weyl group acting on the Coulomb branch as
$\mathbb{R}^r/W(G_r)$, and $\phi_I$ parametrizes the Coulomb branch $\mathbb{R}^r$.
$S_{\rm bkgd}$ is a term which depends only on the background parameters $\beta,\omega_i,m$,
which we shall explain further below. This expression has been proposed with two different
motivations. See \cite{Lockhart:2012vp} for discussions involving topological strings. 
Here, we explain how (\ref{KLSC-index-1}) was proposed from the viewpoint
of 5 dimensional supersymmetric Yang-Mills theory.

First consider the 6d theory on $S^5\times\mathbb{R}$. The partition function (\ref{KLSC-index})
would be computed by a Euclidean 6d theory path integral on $S^5\times S^1$, where
the $S^1$ has circumference $\beta$ and various fields satisfy twisted boundary conditions
due to the extra insertion $-R+a_ij_i+m\cdot\frac{R_1-R_2}{2}$.\footnote{For the convenience of arguments,
we formally assume the existence of a 6d Lagrangian description and the path integral representation of (\ref{KLSC-index}). This is true for the free Abelian $(2,0)$ theory.
For interacting theories, concrete arguments will only rely on the Lagrangian
formulation of the 5d SYM at low energy, which exists.} The twisted boundary conditions
given by $a_ij_i$ can be represented by deforming the background metric of $S^5\times S^1$
in a `complex' manner as follows \cite{Kim:2012qf}:
\begin{eqnarray}\label{KLbackground}
  ds^2(S^5\times S^1)&=&r^2\sum_{i=1}^3
  \left[dn_i^2+n_i^2(d\phi_i+\frac{ia_i}{r}d\tau)^2\right]+d\tau^2\nonumber\\
  &=&r^2\sum_i\left[dn_i^2+n_i^2d\phi_i^2+\alpha^{2}
  (\sum_ja_jn_j^2d\phi_j)^2\right]+\alpha^{-2}\left(d\tau+i\alpha^2
  r\sum_ja_jn_j^2d\phi_j\right)^2\nonumber\\
  &\equiv&g_{\mu\nu}dx^\mu dx^\nu+\alpha^{-2}(d\tau+rC)^2 \ ,
\end{eqnarray}
where $\alpha^{-2}\equiv 1-\sum_jn_j^2a_j^2$ and $C\equiv i\alpha^2\sum_ja_jn_j^2d\phi_j$.
Here $n_i$'s satisfy $n_1^2+n_2^2+n_3^2=1$, and $\tau\sim\tau+\beta$, $\phi\sim\phi_i+2\pi$  
periodicities are assumed.
If one is uncomfortable about the complex metric, one can simply take the chemical potentials
$a_i$'s to be imaginary first, and later continue to real $a_i$'s in the partition function
(\ref{KLSC-index-1}) or 5d SYM. (It will be deforming the action to be complex.) We would
like to understand the partition function (\ref{KLSC-index}) first in the regime $\beta\ll 1$,
in which case one can make the Kaluza-Klein reduction of the 6d theory on a small circle to a
5d SYM on
$S^5$. $\beta=\frac{2\pi r_1}{r}$ in the dimensionless convention is the ratio of the
radii of $S^1$ and $S^5$. In particular, when $\beta\ll 1$, this is identified with the 5d
SYM gauge coupling $g_{YM}^2$ as $\beta=\frac{g_{YM}^2}{2\pi r}$. All terms in
$Z_{\mathbb{R}^4\times T^2}$ appearing in the right hand side can be understood as
non-perturbative instanton corrections for small $\beta$ even from the 5d viewpoint,
as we saw in the section 2.\footnote{Sometimes, (\ref{KLSC-index-1}) makes sense 
even if the small circle reduction does not yield weakly-coupled 5d SYM. For instance, some 
6d $(1,0)$ SCFTs on a circle flow to strongly interacting 5d SCFTs rather than 5d SYMs. 
However, viewing $Z_{\mathbb{R}^4\times T^2}$, $\phi$ as the 6d partition functions and 
6d scalars, (\ref{KLSC-index-1}) still makes sense, although we do not know how to derive it.}

If we Kaluza-Klein reduce the 6d metric on $\tau$ circle, one would naturally expect
to have a supersymmetric Yang-Mills theory on a `squashed' $S^5$ whose metric is given by
$g_{\mu\nu}$ above, also with a background `dilaton' field $\alpha$ and the background
`gravi-photon' field $C$. The last statement can be made more precise by finding
(\ref{KLbackground}) as a 5 dimensional off-shell supergravity background
\cite{Festuccia:2011ws}. We find that this is the case. More precisely, we divide
the construction of 5d SYM on $S^5$ with metric $g_{\mu\nu}$ into two steps. We first
obtaining the vector multiplet part of the action using off-shell supergravity methods, 
which is more cumbersome to achieve in a more conventional method.
We then construct the hypermultiplet part of the action in a more brutal manner. 
The former can be easily done by using the
5d off-shell supergravity of \cite{Kugo:2000af}, which realizes $8$
off-shell SUSY of the background gravity and the dynamical vector multiplets.
Construction of the hypermultiplet part of the action with one off-shell SUSY
closely follows \cite{Hosomichi:2012ek}. The results are
summarized in appendix A.

At this point, let us comment that the metric $g_{\mu\nu}$ of (\ref{KLbackground})
may be just one special way of geometrizing the chemical potentials $\omega_i$. In the
literature, alternative geometric realizations are also discussed, which lead to the
same supersymmetric partition function (\ref{KLSC-index-1})
\cite{Lockhart:2012vp,Qiu:2013aga}.

With the action,  SUSY and notations on the squashed $S^5$ summarized in appendix A,
we can understand the partition function (\ref{KLSC-index-1}) in more detail from 5d SYM.
We first study the classical action at the possible saddle points.
Expanding three $\mathbb{Z}_{\mathbb{R}^4\times T^2}$ factors in the series of
$e^{-\frac{4\pi^2k_i}{\beta\omega_i}}$, with $i=1,2,3$, we find the following factor 
at each value of $k_1,k_2,k_3$ and given $\phi$, 
\begin{equation}\label{KLclassical-action}
  \exp\left[-\frac{1}{\beta}\left(\frac{2\pi^2{\rm tr}(\phi^2)}
  {\omega_1\omega_2\omega_3}+\sum_{i=1}^3\frac{4\pi^2k_i}{\omega_i}\right)\right]\ .
\end{equation}
The exponent can be understood as the action of the following supersymmetric
configurations. The SUSY transformation of the gaugino
$\chi^A$ in the vector multiplet is given by 
\begin{equation}
  \delta\chi^A=\frac{i}{2}(F_{\mu\nu}-\alpha^{-1}\phi V_{\mu\nu})\gamma^{\mu\nu}
  \epsilon^A+\alpha D_\mu(\alpha^{-1}\phi)\gamma^\mu\epsilon^i
  -(D-i\alpha\phi\sigma_3)^A_{\ B}\epsilon^B\ , 
\end{equation}
where $V=dC$. Some off-shell supersymmetric configurations are given by
taking $\chi=0$ and
\begin{equation}\label{KLspecial-saddle}
  F_{\mu\nu}=\phi_0 V_{\mu\nu}\ ,\ \
  \phi=\alpha\phi_0.\ ,\ \ D=i\alpha^2\phi_0\sigma_3
\end{equation}
with constant $\phi_0$, as explained in appendix A. $\phi_0$ can be taken to be
in the Cartan subalgebra, using the global part of the gauge transformation. This is not
the most general supersymmetric configurations. To understand more general
possibilities, we consider
\begin{equation}
  (\delta\chi_A)^\dag(\delta\chi^A)=
  \frac{1}{2f}(\hat{F}_{\mu\nu}\xi^\nu)^2+\frac{1}{2f}
  \left(f\hat{F}_{\mu\nu}-\frac{1}{2}\epsilon_{\mu\nu\alpha\beta\gamma}
  \hat{F}^{\alpha\beta}\xi^\gamma\right)^2+\alpha^2 f\left[D_\mu(\alpha^{-1}\phi)\right]^2
  +f(i\hat{D})^2\ ,
\end{equation}
with $\hat{F}_{\mu\nu}\equiv F_{\mu\nu}-\alpha^{-1}\phi V_{\mu\nu}$,
$\hat{D}\equiv D-i\alpha\phi\sigma_3$. The vector
$\xi=\sum_{i=1}^3\omega_i\frac{\partial}{\partial\phi_i}$ is a Killing spinor 
bilinear: see appendix A. So the following equations define
supersymmetric configurations:
\begin{equation}\label{KLsaddle}
  \hat{F}_{\mu\nu}\xi^\nu=0\ ,\ \
  \hat{F}_{\mu\nu}=\frac{1}{2}\epsilon_{\mu\nu\alpha\beta\gamma}
  \hat{F}^{\alpha\beta}\xi^\gamma/f\ ,\ \ D_\mu(\alpha^{-1}\phi)=0\ ,\ \
  D=i\alpha\phi\sigma_3\ .
\end{equation}
The configuration (\ref{KLspecial-saddle}) is a special solution to these
equations with $\hat{F}_{\mu\nu}=0$. Here, the first two equations are deformations 
of the so-called contact instanton equations \cite{Kallen:2012cs,Harland:2011zs}. Locally, 
the first two equations demand that $\hat{F}_{\mu\nu}$ is orthogonal to the vector $\xi$,
and on the orthogonal 4-plane $\hat{F}_{\mu\nu}$ satisfies the self-duality
condition. Locally, it may look like be a self-duality equation on $\mathbb{R}^4$,
namely an instanton string along $\xi$. But it is highly nontrivial if there would be
globally well defined solutions extending the flat space instanton solutions, or perhaps 
a completely new class of solutions on curved space which do not admit `instanton string' 
picture from the flat space intuition. In particular, for generic $\omega_i$, the vector 
$\xi$ generally does not generate a closed orbit on $T^3$ spanned by the three angles 
$\phi_i$. So naively trying to extend the flat space instanton strings in curved space 
is likely to fail.

We have little idea on the general solutions to the above equations.
There is one class of solutions in which the flat space solutions can be easily
embedded in $S^5$. To understand this, first note that the $\xi$ orbit on $T^3$
closes at $(n_1,n_2,n_3)=(1,0,0)$, $(0,1,0)$ and $(0,0,1)$. This is because $T^3$
degenerates to $S^1$ at these points. So taking the small instanton strings in
flat space (having zero sizes), and letting them wind one of these three circles,
will generate singular configurations with finite action. With nonzero scalar
$\phi=\alpha\phi_0$, we generically should embed these instantons to $U(1)^r\subset G_r$
so that the field strength commutes with $\phi_0$. Let us assign $k_1$, $k_2$, $k_3$
instanton strings to the above three locations. This can be superposed with the special
solutions (\ref{KLspecial-saddle}) since they are all in $U(1)^r$. As explained in
appendix A, plugging in this configuration to the classical action precisely provides
the weight (\ref{KLclassical-action}). This motivates $S_0$ and
$\tau_i=\frac{2\pi i}{\omega_i}$ appearing in the formula (\ref{KLSC-index-1}).
Rigorous treatment is missing at this stage. See also \cite{Qiu:2013aga,Qiu:2014cha}
for detailed discussions on this issue.

With these supersymmetric configurations identified, one should introduce $Q$-exact
deformations which would yield (\ref{KLsaddle}) as saddle point equations, and
then compute the 1-loop determinants. A factorization like \cite{Pestun:2007rz} was
assumed for the 1-loop determinant in \cite{Kim:2012qf} to identify the measure as
given by (\ref{KLSC-index-1}). On $S^4$, the factorization happened due to such a
property of the index theorem which captures the BPS modes contributing to the
determinant, and we expect our factorization in (\ref{KLSC-index-1}) could be derived
by a similar careful treatment.\footnote{Once the factorization is assumed, the
effective $\epsilon_1,\epsilon_2,m$ parameters can be determined by investigating the
coefficients of the bosonic symmetry appearing on the right hand side of $\{Q,S\}$
algebra.} More pragmatically, the factorization has been also shown at a perturbative
level by an independent computation \cite{Qiu:2013aga,Qiu:2014oqa}, which then very
naturally suggests the factorized result (\ref{KLSC-index-1}) at the full non-perturbative
level. The same factorized formula has been obtained by exploring the relation between topological strings and supersymmetric partition functions\cite{Lockhart:2012vp}.   tWith this factorized measure, one should integrate and
sum over the saddle point parameters $\phi_0$, $k_1,k_2,k_3$, which leads to
(\ref{KLSC-index-1}). (We dropped the subscript of $\phi_0$ in \ref{KLSC-index-1}.)

In the next subsection, we shall study the physics of (\ref{KLSC-index-1}) for the
$(2,0)$ theory, in various cases in which (\ref{KLSC-index-1}) can be handled more
concretely.

\subsection{The $(2,0)$ index, $W_N$ characters and Casimir energy}

The index (\ref{KLSC-index-1}) has been studied in more detail for the $(2,0)$ SCFTs,
especially in the $A_{N-1}$ case in which the 5d instanton counting has been best
understood. The technical issue concerning the expression (\ref{KLSC-index-1}) is
that it is given in a `weak coupling' expansion form, taking the form of series expansion
in $\beta$ and $e^{-\frac{4\pi^2}{\beta\omega_i}}$ when $\beta\ll 1$ after $\phi$ integral.
However, the index structure of $Z_{S^5\times S^1}$ will be best visible in the regime
$\beta\gg 1$, as a series expansion in $e^{-\beta}$. At the moment, this re-expansion in
the regime $\beta\gg 1$ has been achieved only in two special cases. One is the 6d
Abelian $(2,0)$ index with all fugacities turned on,
and another is the non-Abelian index with all but one fugacities tuned to special values.
In this subsection, we shall explain these two.\footnote{In principle,
there would be an issue of whether we know the exact form of $Z_{\mathbb{R}^4\times T^2}$
from various expansions only, as explained in
section 2. In all cases in which we made concrete studies, we were able to find exact
expressions which yield the known series expansions in
$\beta^{n}e^{-\frac{4\pi^2k}{\beta}}$ with non-negative integers $n,k$ 
at $\beta\ll 1$}

\hspace*{-.6cm}\textbf{\underline{Abelian $(2,0)$ index}:} We should first explain
what is the virtue of studying the Abelian theory, as the 6d theory is free. In fact
the superconformal index of the free 6d $(2,0)$ tensor multiplet is computed in
\cite{Bhattacharya:2008zy}. In our convention, it is given by
\begin{eqnarray}\label{KLabelian-index}
  Z_{S^5\times S^1}(\beta,m,a_i)&=&e^{-\beta\epsilon_0}
  \exp\left[\sum_{n=1}^\infty\frac{1}{n}f(n\beta,nm,na_i)\right],\\
  f(\beta,m,a_i)&=&\frac{e^{-\frac{3\beta}{2}}(e^{\beta m}+e^{-\beta m})-
  (e^{-\beta(\omega_1+\omega_2)}+e^{-\beta(\omega_2+\omega_3)}+e^{-\beta(\omega_3+\omega_1)})
  +e^{-3\beta}}{(1-e^{-\beta\omega_1})(1-e^{-\beta\omega_2})(1-e^{-\beta\omega_3})}
  \ ,\nonumber
\end{eqnarray}
where we used $\omega_i=1+a_i$. $\epsilon_0$ is the `zero point energy' factor,
which is in general regularization scheme dependent. We shall explain this factor
in more detail below in this subsection. Since we know this (trivial) index concretely,
one might wonder what is the virtue of getting it from (\ref{KLSC-index-1}). The
first reason is simply to check that (\ref{KLSC-index-1}) correctly provides the well
known results. The second reason is to emphasize the precise meaning of the formula 
(\ref{KLSC-index-1}).
The equation (\ref{KLabelian-index}) is given in the form of a series expansion in $e^{-\beta}\ll 1$. 
By expanding
$f(\beta,m,a_i)$ in a series in $e^{-\beta}$, one would obtain an infinite product
expression for $Z_{S^5\times S^1}$ for the Abelian theory:
\begin{eqnarray}\label{KLAbelian-full-index}
  &&\hspace{-1cm}q^{\epsilon_0}\prod_{n_1=0}^\infty\prod_{n_1=0}^\infty\prod_{n_3=0}^\infty
  \frac{(1-q^{2+n_1+n_2+n_3}\zeta_1^{n_1-1}\zeta_2^{n_2}\zeta_3^{n_3})
  (1-q^{2+n_1+n_2+n_3}\zeta_1^{n_1}\zeta_2^{n_2-1}\zeta_3^{n_3})
  (1-q^{2+n_1+n_2+n_3}\zeta_1^{n_1}\zeta_2^{n_2}\zeta_3^{n_3-1})}
  {(1-yq^{\frac{3}{2}+n_1+n_2+n_3}\zeta_1^{n_1}\zeta_2^{n_2}\zeta_3^{n_3})
  (1-y^{-1}q^{\frac{3}{2}+n_1+n_2+n_3}\zeta_1^{n_1}\zeta_2^{n_2}\zeta_3^{n_3})
  (1-q^{3+n_1+n_2+n_3}\zeta_1^{n_1}\zeta_2^{n_2}\zeta_3^{n_3})},\nonumber
\end{eqnarray}
where $q=e^{-\beta}$, $\zeta_i=e^{-\beta a_i}$, $y=e^{-\beta m}$. This is well defined
for small enough $q$.

Now to see if this index is reproduced from (\ref{KLSC-index-1}), we should sum over
all the $q$ series appearing in the $Z_{\mathbb{R}^4\times T^2}$ factors, and make
a `strong coupling' expansion to compare with (\ref{KLAbelian-full-index}). Alternatively,
one can make a `weakly coupled' expansion of 
(\ref{KLabelian-index}) and confirm that we obtain (\ref{KLSC-index-1}). Using suitable
contour integral expression for $\log Z_{S^5\times S^1}$ obtained from (\ref{KLabelian-index})
\cite{Kim:2013nva} or using some properties of the triple sine functions
\cite{Lockhart:2012vp}, one could make an expansion of (\ref{KLabelian-index}) which
is given by 
\begin{eqnarray}
  Z_{S^5\times S^1}\!&\!=\!&\!
  \left[\frac{\beta\omega_1\omega_2\omega_3}{2\pi}\right]^{\frac{1}{2}}\!\!
  e^{-\beta\epsilon_0-\frac{\beta}{24}\left(1+\frac{2a_1a_2a_3+(1-a_1a_2-a_2a_3-a_3a_1)
  (\frac{1}{4}-m^2)+(\frac{1}{4}-m^2)^2}{\omega_1\omega_2\omega_3}\right)}
  e^{\frac{\pi^2(\omega_1^2+\omega_2^2+\omega_3^2-2\omega_1\omega_2-2\omega_2\omega_3
  -\omega_3\omega_1+4m^2)}{24\beta\omega_1\omega_2\omega_3}}\nonumber\\
  &&\cdot Z_{\rm pert}\left(\frac{2\pi i\omega_{21}}{\omega_1},
  \frac{2\pi i\omega_{31}}{\omega_1},2\pi i\left(\frac{m}{\omega_1}+\frac{3}{2}\right)\right)
  Z_{\rm inst}\left(\frac{2\pi i}{\beta\omega_1},\frac{2\pi i\omega_{21}}{\omega_1},
  \frac{2\pi i\omega_{31}}{\omega_1},2\pi i\left(\frac{m}{\omega_1}+\frac{3}{2}\right)\right)
  \nonumber\\
  &&\cdot\left(1,2,3\rightarrow\frac{}{}\!\!2,3,1\right)
  \left(1,2,3\rightarrow\frac{}{}\!\!3,1,2\right)\ ,
\end{eqnarray}
where the last two factors the repetitions of the second line with the $1,2,3$ subscripts
of $\omega_i$ permuted, $Z_{\rm pert}$ is the perturbative $U(1)$ maximal SYM partition
function on the $\Omega$ deformed $\mathbb{R}^4\times S^1$, and $Z_{\rm inst}$ is the `instanton'
part of the $U(1)$ maximal SYM on $\mathbb{R}^4\times S^1$ 
\cite{Nekrasov:2002qd,Nekrasov:2003rj,Iqbal:2008ra}
\begin{equation}\label{KLAbelian-instanton}
  Z_{\rm inst}(\tau,\epsilon_1,\epsilon_2,m_0)=
  \exp\left[\sum_{n=1}^\infty\frac{1}{n}\frac{\sinh\frac{n(m_0+\epsilon_-)}{2}
  \sinh\frac{n(m_0-\epsilon_-)}{2}}
  {\sinh\frac{n\epsilon_1}{2}\sinh\frac{n\epsilon_2}{2}}
  \frac{e^{2\pi in\tau}}{1-e^{2\pi in\tau}}\right]\ ,
\end{equation}
which is identical to (\ref{KLmaximal-instanton-index}) when we expand
(\ref{KLAbelian-instanton}) in $e^{2\pi i\tau}$.
Note that neither $Z_{\rm pert}$, $Z_{\rm inst}$ depends on the $U(1)$ Coulomb VEV $\phi$.
So the first factor $[\frac{\beta\omega_1\omega_2\omega_3}{2\pi}]^{\frac{1}{2}}$ can be 
replaced by 
\begin{equation}
  \left(\frac{\beta\omega_1\omega_2\omega_3}{2\pi}\right)^{\frac{1}{2}}=
  \int_{-\infty}^\infty d\phi
  \exp\left(-\frac{2\pi^2\phi^2}{\beta\omega_1\omega_2\omega_3}\right)\ ,
\end{equation}
which is the Gaussian integral in (\ref{KLSC-index-1}) with the measure $e^{S_0}$.
So the known index (\ref{KLabelian-index}) would be completely agreeing with (\ref{KLSC-index-1})
if we identify $Z_{\mathbb{R}^4\times T^2}=Z_{\rm pert}Z_{\rm inst}$ and if we take
\begin{eqnarray}\label{KLbackground-abelian}
  \epsilon_0&=&-\frac{1}{24}\left[1+\frac{2a_1a_2a_3+(1-a_1a_2-a_2a_3-a_3a_1)
  (\frac{1}{4}-m^2)+(\frac{1}{4}-m^2)^2}{\omega_1\omega_2\omega_3}\right]\ , \nonumber\\
  S_{\rm bkgd}&=&-\frac{\pi^2(\omega_1^2+\omega_2^2+\omega_3^2-2\omega_1\omega_2
  -2\omega_2\omega_3-\omega_3\omega_1+4m^2)}{24\beta\omega_1\omega_2\omega_3}\ .
\end{eqnarray}
We shall explain these three identifications about $\epsilon_0$,  $S_{\rm bkgd}$ and $Z_{\rm inst}$,
  in turn.

Firstly, $\epsilon_0$ is the vacuum `Casimir energy' which we shall explain
later in this subsection. For now, we simply regard (\ref{KLSC-index-1}) as giving
a specific value of the vacuum energy at $\beta\gg 1$ expansion. 
Secondly, $S_{\rm bkgd}$ couples
the parameters $g_{YM}^2,m$ of the theory to the background
parameters of $S^5$, such as $r$, $\omega_i$. In particular, it takes the form of
the leading free energy in the `high temperature' regime $\beta\ll 1$. From the
analysis of one lower dimension on $S^5$, this data cannot be determined in a
self-contained way, and should be given as an input. One can think about it in two
different viewpoints. One may first regard $S_{\rm bkgd}$ as our ignorance, but
demand that we tune it so that the strong coupling expansion of (\ref{KLSC-index-1})
becomes an index. It is an extremely nontrivial request that tuning $S_{\rm bkgd}$
in negative powers of $\beta$ yields an index at $\beta\gg 1$. As the above
results in the Abelian theory shows, it completely fixes $S_{\rm bkgd}$ if one can
freely do both weak and strong coupling expansions. Furthermore, the general
structures of the high temperature asymptotics of the the 6d SCFT index was
proposed in \cite{DiPietro:2014bca}. They only considered the angular momentum
chemical potentials $\omega_i$, with $m=0$, and completely fixed the $\beta,\omega_i$
dependence apart from a few central charge coefficients. Of course their proposal is
consistent with (\ref{KLbackground-abelian}) at $m=0$.

So we finally explain $Z_{\mathbb{R}^4\times T^2}=Z_{\rm pert}Z_{\rm inst}$. At
first sight, it sounds strange that 5d $U(1)$ maximal SYM exhibits such
a nontrivial `instanton' factor $Z_{\rm inst}$, as $U(1)$ adjoint theory looks free.
However, one should understand how Nekrasov's `instanton calculus' yielded
a nontrivial result (\ref{KLAbelian-instanton}). This is because Nekrasov actually
did not work with this free QFT, but worked with a UV completion of it. Namely,
just as we explained in section 2, without any logical reference to 5d SYM,
what we call Nekrasov's `instanton partition function' is a string theory result,
especially for non-renormalizable QFTs. The equation (\ref{KLAbelian-instanton}) gains a solid
meaning as the index for $k$ D0-branes bound to $1$ D4-brane. In our context,
$Z_{\mathbb{R}^4\times T^2}$ is the true 6d CFT observable
computed from the string or M-theory engineering. So although we attempted
to find motivations and supports of the expression (\ref{KLSC-index-1}) from
5d SYM, our true claim is that the integrand $Z_{\mathbb{R}^4\times T^2}$
should naturally be understood as the 6d observable, without necessarily
relying on the UV incomplete 5d SYM. See \cite{Hwang:2014uwa} for more discussions 
on $Z_{\rm inst}$ as a more abstract 6d observable. So understanding that
we should use $Z_{\mathbb{R}^4\times T^2}$ computed from string theory,
the identification $Z_{\mathbb{R}^4\times T^2}=Z_{\rm pert}Z_{\rm inst}$
with (\ref{KLAbelian-instanton}) is justified.

\hspace*{-.6cm}\textbf{\underline{Unrefined non-Abelian $(2,0)$ indices}:}
Now we turn to more interesting non-Abelian indices. Again we shall restrict our
interest to the $(2,0)$ theory here, as we shall crucially use simplifications coming
from extra SUSY when some chemical potentials are tuned. Namely, consider the following
tuning of the $U(1)\subset SO(5)_R$ chemical potential for $\frac{R_1-R_2}{2}$:
\begin{equation}\label{KLunrefine}
  m=\frac{1}{2}-a_3\ .
\end{equation}
The index can then be written as
\begin{equation}
  Z_{S^5\times S^1}(\beta,\frac{1}{2}-a_3,a_i)={\rm Tr}\left[
  (-1)^Fe^{-\beta(E-R_1)}e^{-\beta a_1(j_1-j_3-\frac{R_1-R_2}{2})}
  e^{-\beta(j_2-j_3-\frac{R_1-R_2}{2})}\right]\ .
\end{equation}
Apart from the supercharges $Q\equiv Q^{++}_{---}$, $S\equiv S^{--}_{+++}$ which
commute with this measure by construction, two extra supercharges
$Q^{+-}_{++-}$, $S^{-+}_{--+}$ commute with it at (\ref{KLunrefine}). So the index
exhibits more cancellations of bosons/fermions paired by the extra
supercharges, which will make the index simpler. The SYM on $S^5$ will
also preserve more SUSY. We will show shortly that the equation 
(\ref{KLSC-index-1}) can be exactly computed at this point. We also 
note that further tunings
\begin{equation}\label{KLmaximal-SUSY}
  m=\frac{1}{2}\ ,\ \ a_1=a_2=a_3=0
\end{equation}
will leave only one chemical potential $\beta$, in which case the measure of the index
\begin{equation}\label{KLmaximal-SUSY-index}
  Z_{S^5\times S^1}(\beta,\frac{1}{2},0)={\rm Tr}\left[(-1)^Fe^{-\beta(E-R_1)}\right]
\end{equation}
commutes with $16$ of the $32$ supercharges of the $(2,0)$ theory. Namely, the
following $8$ complex supercharges $Q^{+\pm}_{\pm\pm\pm}$ (with the $\pm$ subscripts
satisfying $\pm\pm\pm=-$) and their conjugates $S^{-\pm}_{\pm\pm\pm}$ (with subscripts
satisfying $\pm\pm\pm=+$) commute with $e^{-\beta(E-R_1)}$. The presence of $16$ SUSY
will have special implication on the index, especially concerning the `zero point energy'
of the vacuum which is captured by the index. Also, one would naturally expect that
the circle reduction of the 6d theory at (\ref{KLmaximal-SUSY}) will yield a maximal
SYM which actually preserves $16$ supercharges. This is indeed the case
\cite{Kim:2012ava}. See also \cite{Minahan:2013jwa}.

To understand the simplification at the level of the formula (\ref{KLSC-index-1}),
we first study how the $\Omega$ background parameters and the mass parameters simplify. 
In the notation of section 2, the
effective $\Omega$ parameters $\epsilon_1,\epsilon_2$ and the mass $m_0$ in the three 
$Z_{\mathbb{R}^4\times T^2}$ factors are given by
\begin{eqnarray}
  \frac{1}{2\pi i}(\epsilon_1,\epsilon_2,m_0)&=&\left(\frac{\omega_2-\omega_1}{\omega_1},
  \frac{\omega_3-\omega_1}{\omega_1},\frac{m}{\omega_1}+\frac{3}{2}\right)\sim
  \left(\frac{\omega_2-\omega_1}{\omega_1},
  \frac{\omega_3-\omega_1}{\omega_1},\frac{m}{\omega_1}-\frac{1}{2}\right)\ :\ \textrm{1st}\nonumber\\
  &&\left(\frac{\omega_3-\omega_2}{\omega_2},
  \frac{\omega_1-\omega_2}{\omega_2},\frac{m}{\omega_2}-\frac{1}{2}\right)\ :\ \textrm{2nd}\nonumber\\
  &&\left(\frac{\omega_1-\omega_3}{\omega_3},
  \frac{\omega_2-\omega_3}{\omega_3},\frac{m}{\omega_3}-\frac{1}{2}\right)\ :\ \textrm{3rd}\ .
\end{eqnarray}
We use $m_0$ for the effective mass parameter on $\mathbb{R}^4\times T^2$, to avoid confusions
with the actual mass parameter of 5d SYM on $S^5$, or the chemical potential $\beta m$ of the
6d index. Also, we used the fact that all parameters $\epsilon_1,\epsilon_2,m_0$ are periodic 
variables in $2\pi i $ shifts. So at (\ref{KLunrefine}), one finds 
\begin{eqnarray}
  \frac{1}{2\pi i}(\epsilon_1,\epsilon_2,m_0)&=&
  \left(\frac{\omega_2-\omega_1}{\omega_1},
  \frac{\omega_3-\omega_1}{\omega_1},\frac{\omega_2-\omega_3}{2\omega_1}\right)
  \ :\ \textrm{1st}\nonumber\\
  &&\left(\frac{\omega_3-\omega_2}{\omega_2},
  \frac{\omega_1-\omega_2}{\omega_2},\frac{\omega_1-\omega_3}{2\omega_2}\right)
  \ :\ \textrm{2nd}\nonumber\\
  &&\left(\frac{\omega_1-\omega_3}{\omega_3},
  \frac{\omega_2-\omega_3}{\omega_3},\frac{\omega_1+\omega_2-2\omega_3}{2\omega_3}\right)
  \ :\ \textrm{3rd}\ .
\end{eqnarray}
Defining $\epsilon_\pm\equiv\frac{\epsilon_1\pm\epsilon_2}{2}$,
we find that these effective parameters satisfy $m_0=\epsilon_-$ in the first factor,
$m_0=-\epsilon_-$ in the second factor, and $m=\epsilon_+$ in the third factor.

Let us explain that the partition function $Z_{\mathbb{R}^4\times T^2}$ simplifies
in all the three factors. Note that
$Z_{\mathbb{R}^4\times T^2}(q,v,\epsilon_{1,2},m_0)$ takes the following form:
\begin{eqnarray}\label{KLcoulomb-recall}
  Z_{\mathbb{R}^4\times T^2}&=&Z_{\rm pert}(v,\epsilon_{1,2},m_0)Z_{\rm inst}(q,v,\epsilon_{1,2},m_0)\ , 
  \\
  Z_{\rm pert}&=& \widetilde{PE}\left[\frac{1}{2}\ \frac{\sin\frac{m_0+\epsilon_+}{2}\frac{m_0-\epsilon_+}{2}}
  {\sin\frac{\epsilon_1}{2} \sin\frac{\epsilon_2}{2}}\chi_{\rm adj}(v)\right]\ , \nonumber
\end{eqnarray}
where $Z_{\rm inst}$ is given in section 2, and 
\begin{equation}
  \chi_{\rm adj}(v)=\sum_{\alpha\in{\rm adj}(G)}e^{\alpha(v)}\ .
\end{equation}
$\widetilde{PE}$ is defined by expanding the function in $\widetilde{PE}[\cdots]$ in $e^{-\epsilon_{1,2}}$, $e^{-m_0}$, $e^{-\alpha(v_i)}$, and imposing
\begin{equation}
  \widetilde{PE}[ne^{-x}]=\left[2\sinh\frac{x}{2}\right]^n=
  \left[\frac{e^{-\frac{x}{2}}}{1-e^{-x}}\right]^n\ ,\ \
  \widetilde{PE}[f+g]=\widetilde{PE}[f]\widetilde{PE}[g]\ .
\end{equation}
As for $Z_{\rm inst}$, the $U(N)$ result is known well in the series expansion in $q$,
as explained in section 2. For $D_N$ cases, the $SO(2N)$ partition function is in principle
computable from \cite{Hwang:2014uwa}, but not very much have been done in detail so far.
For $E_N$, almost nothing is known, although we shall say something about it below. Here we would
like to emphasize the general structure of $Z_{\rm pert}$ and $Z_{\rm inst}$. 
Since $Z_{\rm pert}$ and
$Z_{\rm inst}$ count BPS particles on $\mathbb{R}^{4,1}$ in the Coulomb phase, they carry
universal prefactors from their center-of-mass supermultiplets. In particular, since
perturbative particles and instantons preserve different $8$ supercharges among the full $16$ as massive vector and tensor multiplets, respectively, 
the prefactors appearing in the two parts are different. It is easy to
check \cite{Kim:2011mv} that
\begin{equation}\label{KLindex-com}
  Z_{\rm pert}=\widetilde{PE}\left[I_+(\epsilon_{1,2},m_0)(\cdots)\right]\ ,\ \
  Z_{\rm inst}=\widetilde{PE}\left[I_-(\epsilon_{1,2},m_0)(\cdots)\right]\ .
\end{equation}
$(\cdots)$ are the contributions from internal degrees of freedom of the BPS states,
which are regular in $\epsilon_1=\epsilon_2=m_0=0$ the limit, and
\begin{equation}\label{KLcom}
  I_\pm(\epsilon_{1,2},m_0)\equiv\frac{\sin\frac{m_0+\epsilon_\pm}{2}\sin\frac{m_0-\epsilon_\pm}{2}}
  {\sin\frac{\epsilon_1}{2}\sin\frac{\epsilon_2}{2}}\ .
  \end{equation}
For $Z_{\rm pert}$, this structure is already manifest in (\ref{KLcoulomb-recall}).

Firstly, at $m_0=\pm\epsilon_-$, one finds from (\ref{KLcom}) and (\ref{KLindex-com}) that
$I_-(\epsilon_{1,2},\pm\epsilon_-)=0$ and $Z_{\rm inst}=1$. Therefore, in the unrefined
limit (\ref{KLunrefine}), one finds that the first and second $Z_{\mathbb{R}^4\times T^2}$
factors in (\ref{KLSC-index-1}) reduces to the perturbative contributions at this point.
Note also from (\ref{KLcom}) that $I_+(\epsilon_{1,2},\pm\epsilon_-)=-1$. So applying
this to (\ref{KLcoulomb-recall}), one obtains
\begin{equation}
  Z_{\mathbb{R}^4\times T^2}\rightarrow \widetilde{PE}\left[-\frac{1}{2}\chi_{\rm adj}(v)\right]
\end{equation}
at $m_0=\pm\epsilon_-$. So applying this to the first and second factors of (\ref{KLSC-index-1}),
one obtains
\begin{equation}\label{KL1st-2nd-unrefined}
  Z_{\mathbb{R}^4\times T^2}^{(1)}Z_{\mathbb{R}^4\times T^2}^{(2)}
  \rightarrow \widetilde{PE}\left[-\frac{1}{2}(\chi_{\rm adj}(v/\omega_1)+
  \chi_{\rm adj}(v/\omega_2)\right]=\prod_{\alpha>0}2\sinh\frac{\alpha(v)}{\omega_1}
  \cdot 2\sinh\frac{\alpha(v)}{\omega_2}\ ,
\end{equation}
where the product is over positive roots of $G$,
up to a possible overall sign on which we are not very careful. 
We then turn to $Z^{(3)}_{\mathbb{R}^4\times T^2}$
at $m_0=\epsilon_+$. From (\ref{KLindex-com}) or (\ref{KLcoulomb-recall}), it is obvious that
$Z_{\rm pert}=1$ at $m_0=\epsilon_+$, since $I_+=0$. So $Z_{\mathbb{R}^4\times T^2}$ acquires
$Z_{\rm inst}$ contribution only at $m=\epsilon_+$. For $U(N)$, one can easily show from
the $U(N)$ instanton partition function of section 2  that
\begin{equation}
  Z_{\rm inst}(\beta,\epsilon_{1,2},m_0=\pm\epsilon_+)=\frac{1}{\eta(\tau)^N}=
  e^{-\frac{\pi Ni\tau}{12}}\prod_{n=1}^\infty\frac{1}{(1-e^{2\pi ni\tau})^N}\ .
\end{equation}
where $\tau=\frac{2\pi i}{\beta}$. Here, we have included the extra factor of
\begin{equation}\label{KLbackground-unrefined}
  e^{-S_{\rm bkgd}}=e^{-\frac{\pi Ni\tau}{12}}=e^{\frac{\pi^2 N}{6\beta}},
\end{equation}
which will be justified below. More generally, we shall present below
nontrivial evidences that
\begin{equation}\label{KLinstanton-ADE}
  e^{-S_{\rm bkgd}}Z_{\rm inst}(\beta,\epsilon_{1,2},m_0=\pm\epsilon_+)=\frac{1}{\eta(\tau)^N}
\end{equation}
for all $U(N)$, $D_N=SO(2N)$, $E_N$ groups. If one wishes to consider the interacting
$A_{N-1}$ part only, instead of $U(N)$, one simply takes $Z_{\rm inst}=\eta(\tau)^{-(N-1)}$
by dropping an overall $U(1)$ factor $Z_{\rm inst}^{U(1)}=\eta(\tau)^{-1}$. So
$e^{-S_{\rm bkgd}}Z_{\mathbb{R}^4\times T^2}^{(3)}$ simplifies  to
\begin{equation}\label{KL3rd-unrefined}
  e^{-S_{\rm bkgd}}Z^{(3)}_{\mathbb{R}^4\times T^2}\rightarrow\frac{1}{\eta(\tau/\omega_3)^N}
  =\frac{1}{\eta(\frac{2\pi i}{\beta\omega_3})^N}
\end{equation}
in the limit (\ref{KLunrefine}), for all $A_N$, $D_N$, $E_N$ series.

With all the simplifications (\ref{KL1st-2nd-unrefined}), (\ref{KL3rd-unrefined}), the partition
function (\ref{KLSC-index-1}) on $S^5$ reduces to
\begin{equation}\label{KLunrefined-collect}
  Z_{S^5\times S^1}(\beta,m=\frac{1}{2}-a_3,a_i)=\frac{1}{\eta(\frac{2\pi i}{\beta\omega_3})^N}
  \cdot\frac{1}{W(G_N)}\int\prod_{I=1}^Nd\phi_I
  e^{-\frac{2\pi^2{\rm tr}(\phi^2)}{\beta\omega_1\omega_2\omega_3}}\prod_{\alpha>0}
  2\sinh\frac{\alpha(\phi)}{\omega_1}\cdot 2\sinh\frac{\alpha(\phi)}{\omega_2}\ .
\end{equation}
The integral over $\phi_I$ is simply a Gaussian integral. The result of the integral is
\begin{equation}\label{KLunrefined-integral}
  Z_{S^5\times S^1}(\beta,m=\frac{1}{2}-a_3,a_i)=
  \left(\frac{\beta\omega_3}{2\pi}\right)^{\frac{N}{2}}\frac{1}{\eta(\frac{2\pi i}{\beta\omega_3})^N}
  \ e^{\frac{\beta c_2|G|}{24}\omega_3
  \left(\frac{\omega_1}{\omega_2}+\frac{\omega_2}{\omega_1}\right)}
  \prod_{\alpha>0}2\sinh\left(\beta\omega_3\frac{\alpha\cdot\rho}{2}\right)\ , 
\end{equation}
where $\rho$ is the Weyl vector. Since $\eta(\tau)$ is a modular form, its expansion in
the $\beta\gg 1$ regime is easy to understand. The result is given by
\begin{eqnarray}
  Z_{S^5\times S^1}(\beta,\frac{1}{2}-a_3,a_i)&=&
  e^{\beta\frac{c_2|G|}{24}\frac{\omega_3}{\omega_1\omega_2}\left(\omega_1+\omega_2\right)^2}
  \prod_{\alpha>0}(1-e^{-\beta\omega_3\alpha\cdot\rho})\cdot\frac{1}{\eta(\frac{i\beta\omega_3}{2\pi})^N}\\
  &=&e^{\beta\frac{c_2|G|}{24}\frac{\omega_3}{\omega_1\omega_2}\left(\omega_1+\omega_2\right)^2
  +\frac{N\beta\omega_3}{24}}
  \prod_{\alpha>0}(1-e^{-\beta\omega_3\alpha\cdot\rho})\cdot\prod_{n=1}^\infty
  \frac{1}{(1-e^{-n\beta\omega_3})^N}\ .\nonumber
\end{eqnarray}
After a little computation  for the cases $G=U(N)$, $D_N$, $E_N$ (the reference \cite{Kim:2012qf} for the $A$ and $D$ cases),
one obtains
\begin{eqnarray}\label{KLunrefined-final}
  Z_{S^5\times S^1}(\beta,\frac{1}{2}-a_3,a_i)&=&e^{\beta\frac{c_2|G|}{24}
  \frac{\omega_3}{\omega_1\omega_2}\left(\omega_1+\omega_2\right)^2+\frac{N\beta\omega_3}{24}}
  \prod_{s=0}^\infty\prod_{d=\textrm{deg}[C(G)]}\frac{1}{1-e^{-\beta\omega_3(d+s)}}\ ,
\end{eqnarray}
where $d$ runs over the degrees of the possible Casimir operators $C(G)$ of the group $G$.
More concretely, the degrees of the Casimir operators are
\begin{eqnarray}\label{KLdimension-casimir}
  U(N)&:&1,2,\cdots,N \\
  SO(2N)&:&2,4,\cdots,2N-2\ \textrm{and}\ N\nonumber\\
  E_6&:&2,5,6,8,9,12\nonumber\\
  E_7&:&2,6,8,10,12,14,18\nonumber\\
  E_8&:&2,8,12,14,18,20,24,30\nonumber
\end{eqnarray}
for all ADE groups. We shall shortly give physical interpretations of these results.

Before proceeding to the interpretation of the result, we emphasize that the
expression (\ref{KLSC-index-1}) obtained at $\beta\ll 1$ successfully becomes an
index (or more generally, partition function which counts states) in the $\beta\gg 1$ regime,
meaning that an expansion in $e^{-\beta}\ll 1$ has
integer coefficients only. At this stage, we can justify the choice of
$S_{\rm bkgd}=-\frac{\pi^2N}{6\beta}$. Just as in the Abelian case, we had to add this
part by hand, as our supports on (\ref{KLSC-index-1}) came from one
lower dimension in the high temperature regime. Namely, the leading singular behaviors of
the free energy (coming in negative powers of $\beta$) have to be inputs in this approach.
This input is all encoded in $S_{\rm bkgd}$, in the form of the couplings of the parameters
of the theory to the background gravity fields. As in the Abelian case, $S_{\rm bkgd}$
in negative powers of $\beta$ is uniquely fixed by demanding the full quantity to be
an index at $\beta\gg 1$. The structure of such couplings $S_{\rm bkgd}$ has been
explored in \cite{DiPietro:2014bca} in the case of $S^3\times S^1$ using 3 dimensional 
supergravity, and similar studies are made on $S^5\times S^1$. In particular, the absence 
of a term proportional to $\beta^{-3}$ in our $S_{\rm bkgd}$ is consistent with what 
\cite{DiPietro:2014bca} proposes for the $(2,0)$ theory.

Also note that the choice (\ref{KLinstanton-ADE}) for all ADE group is consistent
with the requirement to have an index at $\beta\gg 1$, because the modular transformation
of $\eta^{-N}$ to go to the $\beta\gg 1$ regime exactly absorbs the factor
$\left(\frac{\beta\omega_3}{2\pi}\right)^{\frac{N}{2}}$ in (\ref{KLunrefined-integral}), which
could have obstructed the index interpretation. Below we shall provide more
support of our choice (\ref{KLinstanton-ADE}) for DE groups.

We now study the physics of (\ref{KLunrefined-final}).
We first consider the `spectrum' part of this index,
\begin{equation}\label{KLunrefined-spectrum}
  \prod_{s=0}^\infty\prod_{d=\textrm{deg}[C(G)]}\frac{1}{1-q^{d+s}}
  =PE\left[\frac{\sum_{d=\textrm{deg}[C(G)]}q^d}{1-q}\right]\ ,
\end{equation}
where we defined $q\equiv e^{-\beta\omega_3}$, and $PE$ here is defined in a more
standard manner, $PE[f(x)]\equiv\exp\left[\sum_{n=1}^\infty\frac{1}{n}
f(x^n)\right]$, without including the zero point energy factors. All coefficients
of this index in $q$ expansion has positive coefficients, implying the possibility
that this could actually be a partition function counting bosonic states/operators only.
This is independently supported by other studies on the 6d $(2,0)$ theory
\cite{Beem:2014kka},
which identified a closed 2d  bosonic chiral subsector of the operator product expansions of local
operators. 

To give a more intuitive feelings on (\ref{KLunrefined-spectrum}), we shall first explain 
an analogous situation in the 4d $\mathcal{N}=4$ Yang-Mills theory with ADE gauge groups, 
in which (\ref{KLunrefined-spectrum}) also emerges as the partition function of a class of
BPS operators. In the 4d SYM, we are interested
in gauge invariant BPS operators in the weakly coupled theory, consisting of one complex
scalar $\Phi$ and one of the two holomorphic derivatives on $\mathbb{R}^4$,
which we call $\partial$. The spectrum of these operators in the weakly coupled regime is
worked out in \cite{Grant:2008sk}. In particular, we are interested in local operators
$\mathcal{O}$
which are annihilated by a specific $Q$, $Q\mathcal{O}=0$, with the equivalence relation
$\mathcal{O}\sim\mathcal{O}+Q\lambda$, so we are interested in the cohomology elements.
It was shown that the coholomology elements can be constructed using the $\Phi$ letters
with $\partial$ derivatives only. The cohomology elements can be constructed by
multiplying elements of the form
\begin{equation}\label{KLchiral-basis}
  \partial^{s}f(\Phi)\ ,
\end{equation}
where $f(\Phi)$ is any gauge invariant expression for the matrix $\Phi$, and then
linearly superposing all possible operators constructed this way. So the question is
to find the independent `generators' taking the form of (\ref{KLchiral-basis}). Note
that if $f$ satisfies $f(\Phi)=g(\Phi)h(\Phi)$ or $f(\Phi)=g(\Phi)+h(\Phi)$ with
other gauge invariant expressions $g(\Phi),h(\Phi)$, then (\ref{KLchiral-basis}) is
no longer an independent generator. With these considerations, if one takes
$f(\Phi)$ to be all possible independent Casimir operators of the gauge group,
then (\ref{KLchiral-basis}) forms the complete set of generators of the cohomology.
The dimension of $\Phi$ is $1$, so the dimension of the operator $f(\Phi)$ is the
degree of the Casimir operator. So for instance, for ADE gauge groups, this leads to
the scale dimension spectrum (\ref{KLdimension-casimir}) of the $f(\Phi)$ appearing in the
generator (\ref{KLchiral-basis}). For instance, the generators for $U(N)$ are
$f(\Phi)={\rm tr}(\Phi^n)$ with $n=1,\cdots,N$. The generators for $SU(N)$ also takes
the same form, with $n=2,\cdots,N$. The generators for $SO(2N)$ are
$f(\Phi)={\rm tr}(\Phi^2),\cdots,{\rm tr}(\Phi^{2N-2})$ and
${\rm Pf}(\Phi)=\sqrt{\det\Phi}$. Thus the partition function for these generators,
where the letters with scale dimension $\Delta$ are weighted by $q^\Delta$,
is given by
\begin{equation}
  z(q)_{\rm letter}=\sum_{s=0}^\infty\sum_{d\in{\rm deg}[C(G)]}q^{d+s}=
  \frac{\sum_{d\in{\rm deg}[C(G)]}q^{d}}{1-q}\ ,
\end{equation}
where $\Delta[\Phi]=\Delta[\partial]=1$. Now the full set of the cohomology is obtained
by forming the Fock space of the generators (\ref{KLchiral-basis}), and the partition function
over this space is given by (\ref{KLunrefined-spectrum}).

To summarize, from 4d maximal SYM, we obtained the same partition function as 
what we got for the 6d $(2,0)$ theory. This is not strange.
For instance, had we been counting gauge invariant operators made of scalar $\Phi$ only
without any derivatives, this would have given us the half-BPS states whose partition
function is given by $PE[\sum_{d\in{\rm deg}[C(G)]}q^d]$. The half-BPS partition function
is known to be universal in all maximal superconformal field theories, in 3,4,6 dimensions.
There is also an explanation of this universality, by quantizing and counting the states of
half-BPS giant gravitons in the AdS duals \cite{Mandal:2006tk,Bhattacharyya:2007sa}.
Even after including one derivative, one can follow the D3-brane giant graviton counting
of the partition function (\ref{KLunrefined-spectrum}) in $AdS_5\times S^5$
\cite{Ashok:2008fa,Grant:2008sk}, to quantize and count the M5-brane giant gravitons on
$AdS_7\times S^4$ to obtain the same partition
function (at least for the A and D series). At this point, let us
mention that the large $N$ limits of (\ref{KLunrefined-spectrum}) for $U(N)$ and $SO(2N)$
completely agree with the supergravity indices on $AdS_7\times S^4$ and
$AdS_7\times S^4/\mathbb{Z}_2$, respectively \cite{Kim:2012ava,Kim:2012qf}.
It is also reassuring to find that the chiral algebra arguments of \cite{Beem:2014kka}
naturally suggest the same partition function (\ref{KLunrefined-spectrum}), for all
ADE cases. So turning the logic around, the natural result (\ref{KLunrefined-spectrum})
also supports our conjecture (\ref{KLinstanton-ADE}) on the instanton correction for
the gauge groups DE. (The case with $SO(2N)$ may be derivable with the results of
\cite{Hwang:2014uwa}.)

The partition function (\ref{KLunrefined-spectrum}), with $d$ running over the
degrees of the Casimir operators of $G=SU(N)$, $SO(2N)$, $E_N$, is known to be
the vacuum character of the $W_G$ algebra. The appearance of the $W_G$
algebra in the superconformal index, and more generally in the chiral subsector
of the 6d BPS operators, was asserted in \cite{Beem:2014kka} to be closely related to
the appearance of the 2d Toda theories in the AGT correspondence
\cite{Alday:2009aq,Wyllard:2009hg}. In fact the appearance of $W_G$ algebra and
relation to the AGT relation are further supported recently, by considering the
superconformal index with various defect operators \cite{Bullimore:2014upa}. Namely,
for the $A_{N-1}$ theories, insertions of various
dimension 2 and/or 4 defect operators to the unrefined index yielded the characters
of various degenerate and semi-degenerate representations of $W_N$ algebra, and
also the characters of the so-called $W_N^\rho$ representations when the dimension
4 operator is wrapped over the 2d plane where the chiral operators live. This appears to
be very concrete supports of the predictions of the AGT correspondence \cite{Alday:2009fs}
from the 6d index.

Finally, let us explain the prefactor of (\ref{KLunrefined-final}), which takes the
form of $e^{-\beta(\epsilon_0)_{\rm SUSY}}$ with
\begin{equation}\label{KLcasimir}
  (\epsilon_0)_{\rm SUSY}=-\frac{c_2|G|}{24}\frac{\omega_3}{\omega_1\omega_2}
  \left(\omega_1+\omega_2\right)^2-\frac{N\omega_3}{24}\ .
\end{equation}
This formally takes the form of the `vacuum energy' as it is conjugate to the chemical
potential $\beta$ in the index. However, one needs to understand vacuum energies with
care. As is obvious already in the free quantum field theory, vacuum energy is the summation
of zero point energies of infinitely many harmonic oscillators, which is formally
divergent. It is a quantity that has to be carefully defined and computed with
regularization/renormalization. Since the regularization and renormalization are
constrained by symmetry, it will be simplest to start the discussion with the special
case (\ref{KLmaximal-SUSY}), (\ref{KLmaximal-SUSY-index}) of our index. In this case,
(\ref{KLcasimir}) simplifies to
\begin{equation}\label{KLcasimir-maximal}
  (\epsilon_0)_{\rm SUSY}=-\frac{c_2|G|}{6}-\frac{N}{24}\ .
\end{equation}
More general cases will be commented on later.

Since $\beta$ is conjugate to $E-R_1$ in (\ref{KLmaximal-SUSY-index}), the formal definition 
of $(\epsilon_0)_{\rm SUSY}$ is given by
the `expectation value' of the charge $E-R_1$ for the vacuum on $S^5\times\mathbb{R}$,
\begin{equation}\label{KLvacuum-energy}
  \langle E-R_1\rangle=-\left.\frac{\partial}{\partial\beta}\log Z_{S^5\times S^1}
  \right|_{\beta\rightarrow\infty}\ .
\end{equation}
This quantity has to be carefully defined. To concretely illustrate
the subtleties, it will be illustrative to consider the
free $(2,0)$, consisting of an Abelian tensor multiplet.
Then, (\ref{KLvacuum-energy}) is given by the collection of the zero point
values of $E-R_1$ for the free particle oscillators:
\begin{equation}\label{KLfree-casimir-index}
  (\epsilon_0)_{\rm SUSY}={\rm tr}\left[(-1)^F\frac{E-R_1}{2}\right]=
  \sum_{\rm bosonic\ modes}\frac{E-R_1}{2}
  -\sum_{\rm fermionic\ modes}\frac{E-R_1}{2}\ .
\end{equation}
The trace is over the infinitely many free particle modes, and $E,R_1$ appearing in 
the sum are the values of $E,R_1$ carried by modes. This is similar to
the ordinary Casimir energy defined by
\begin{equation}\label{KLfree-casimir-proper}
  \epsilon_0\equiv{\rm tr}\left[(-1)^F\frac{E}{2}\right]=\sum_{\rm bosonic\ modes}\frac{E}{2}
  -\sum_{\rm fermionic\ modes}\frac{E}{2}\ ,
\end{equation}
which appears when one computes the partition function of a QFT on $S^n\times\mathbb{R}$
with inverse-temperature $\beta$ conjugate to the energy $E$ \cite{Aharony:2003sx}.
Both expressions are formal, and should be supplemented by a suitable regularization
of the infinite sums. As in \cite{Aharony:2003sx} for the latter quantity, one can use
the charges carried by the summed-over states to provide regularizations.
The charges that can be used in the regulator are constrained by the symmetries of
the problem under considerations, which are different between
(\ref{KLfree-casimir-index}) and (\ref{KLfree-casimir-proper}). The latter is what is
normally called the Casimir energy. Let us call $(\epsilon_0)_{\rm SUSY}$
the supersymmetric Casimir energy \cite{Cassani:2014zwa}.

For (\ref{KLfree-casimir-proper}), the only charge that one can use to regularize the sum
is energy $E$ \cite{Aharony:2003sx}. This is because the symmetry of the path integral
for the partition function on $S^n\times S^1$ includes all the internal symmetry of the
theory, together with the rotation symmetry on $S^n$. Firstly, non-Abelian rotation
symmetries forbid nonzero vacuum expectation values of angular momenta on $S^n$. Also,
there are no sources which will give nonzero values for the internal charges: its
expectation value is zero either if the internal symmetry is non-Abelian, or
if there are sign flip symmetries of the Abelian internal symmetries. On the other hand,
energy $E$ can be used in the regulator function.
The remaining procedure of properly defining (\ref{KLfree-casimir-proper}) is explained
in \cite{Aharony:2003sx}. One introduces a regulator function $f(E/\Lambda)$ with
a UV cut-off $\Lambda$ (to be sent back to infinity at the final stage) which satisfies
the properties $f(0)=1$, $f(\infty)=0$ and is sufficiently flat at $E/\Lambda=0$: $f^\prime(0)=0$,
$f^{\prime\prime}(0)=0$, etc. The rigorous definition replacing
(\ref{KLfree-casimir-proper}) is given by
\begin{equation}
  {\rm tr}\left[(-1)^F\frac{E}{2}f(E/\Lambda)\right]\ .
\end{equation}
When energy level $E$ has an integer-spacing, $E=\frac{m}{R}$ with $m=1,2,3,\cdots$,
and the degeneracy for given $m$ is a polynomial of $m$ (as in \cite{Aharony:2003sx}),
one can show that this definition is the same as
\begin{equation}\label{KLreg-casimir-proper}
  {\rm tr}\left[(-1)^F\frac{E}{2}e^{-\beta^\prime E}\right]
  =-\frac{1}{2}\frac{d}{d\beta^\prime}{\rm tr}\left[(-1)^Fe^{-\beta^\prime E}\right]\ ,
\end{equation}
where small $\beta^\prime$ is the regulator here. We shall use the latter regulator 
in our discussions.

On the other hand, the correct regularization of (\ref{KLfree-casimir-index}) is
constrained by different symmetries. At $m=\pm\frac{1}{2}$, $a_i=0$,
the symmetry of the path integral is $SU(4|2)$ subgroup of $OSp(8^\ast|4)$,
containing $16$ supercharges. For instance, this is visible on the 5d SYM on $S^5$
\cite{Kim:2012ava}, and is also manifest in (\ref{KLmaximal-SUSY-index}).
This subgroup is defined by elements of $OSp(8^\ast|4)$ which commute with $E-R_1$.
So to respect the $SU(4|2)$ symmetry, only $E-R_1$ can be used to regularize the sum
(\ref{KLfree-casimir-index}). So this sum can be
regularized as ${\rm tr}\left[(-1)^F\frac{E-R_1}{2}f(\frac{E-R_1}{\Lambda})\right]$,
or equivalently as
\begin{equation}\label{KLreg-casimir-index}
  {\rm tr}\left[(-1)^F\frac{E-R_1}{2}e^{-\beta^\prime(E-R_1)}\right]=
  -\frac{1}{2}\frac{d}{d\beta^\prime}{\rm tr}\left[(-1)^Fe^{-\beta^\prime(E-R_1)}\right]\ .
\end{equation}
The quantities
\begin{equation}
  f_{\rm SUSY}(\beta^\prime)={\rm tr}\left[(-1)^F e^{-\beta^\prime(E-R_1)}\right]\ ,\ \
  f(\beta^\prime)={\rm tr}\left[(-1)^F e^{-\beta^\prime E}\right]
\end{equation}
are computed in \cite{Bhattacharya:2008zy,Kim:2013nva} for the free 
$(2,0)$ tensor multiplet, given by
\begin{eqnarray}
  f(x)&=&\frac{5x^2(1-x^2)-16x^{\frac{5}{2}}(1-x)+(10x^3-15x^4+6x^5-x^6)}{(1-x)^6}\ , \nonumber\\
  f_{\rm SUSY}(x)&=&\frac{x}{1-x}\ ,
\end{eqnarray}
where $x\equiv e^{-\frac{\beta^\prime}{r}}$ with the $S^5$ radius $r$.
From these expressions, one obtains
\begin{equation}\label{KLcasimir-reg}
  -\frac{1}{2}\frac{d}{d\beta^\prime}f(x)
  =\frac{5r}{16(\beta^\prime)^2}-\frac{25}{384r}+r^{-3}\mathcal{O}(\beta^\prime)^2
\end{equation}
and
\begin{equation}\label{KLSUSY-casimir-reg}
  -\frac{1}{2}\frac{d}{d\beta^\prime}f_{\rm SUSY}(x)=
  \frac{r}{2(\beta^\prime)^2}-\frac{1}{24r}+r^{-3}\mathcal{O}(\beta^\prime)^2\ ,
\end{equation}
as $\beta^\prime\rightarrow 0$. As explained in \cite{Aharony:2003sx}, the first term
$\frac{5r}{16(\beta^\prime)^2}\sim\frac{5}{16}r\Lambda^2$ of (\ref{KLcasimir-reg}) should
be canceled by a counterterm.
This is because the vacuum value of $E$ has to be zero in the flat
space limit $r\rightarrow\infty$ from the conformal symmetry. A counterterm of the form
$\Lambda^2\int_{S^5\times S^1} d^6x\sqrt{g}\ R^2$
or $(\beta^\prime)^{-2}\int_{S^5\times S^1} d^6x\sqrt{g}\ R^2$ can cancel this divergence.
Similarly, the first term of (\ref{KLSUSY-casimir-reg}) has to be canceled by a
counterterm of the same form. This is
because the vacuum value of $E-R_1$ has to vanish in the flat space limit, required
by the superconformal symmetry. After these subtractions and removing
the regulator $\beta^\prime\rightarrow 0$, one obtains
\begin{equation}\label{KLcasimir-abelian}
  \epsilon_0=-\frac{25}{384r}\ ,\ \ (\epsilon_0)_{\rm SUSY}=-\frac{1}{24r}\ .
\end{equation}
So although conceptually closely related, the two quantities are
different observables. At least with the Abelian example above,
we hope that we clearly illustrated the difference.

Considering that our $S^5$ partition function is constrained by $SU(4|2)$ SUSY
in the path integral, it is very natural to expect that $(\epsilon_0)_{\rm SUSY}$ of
(\ref{KLcasimir-maximal}) is the supersymmetric Casimir energy. Note also that,
$(\epsilon_0)_{SUSY}=-\frac{1}{24r}$ computed above for the free 6d theory agrees
with the zero point energy (\ref{KLcasimir-maximal}) computed from the $S^5$ partition
function at $N=1$, which concretely supports this natural expectation.
(Note that $c_2|G|\equiv f^{abc}f^{abc}=0$ for Abelian gauge group, and also
that we absorbed the factor $\frac{1}{r}$ into $\beta$.) Even in the non-Abelian case, we think that (\ref{KLcasimir-maximal})
should be the supersymmetric Casimir energy $(\epsilon_0)_{\rm SUSY}$, and
not $\epsilon_0$.

The more conventional Casimir energy $\epsilon_0$ in the large $N$ limit has been
computed in the $AdS_7\times S^4$ gravity dual in the literature. The result is
given by \cite{Awad:2000aj,Kallen:2012zn}
\begin{equation}
  \epsilon_0=-\frac{5N^3}{24r}\ ,
\end{equation}
while from the 5d maximal SYM with $U(N)$ gauge group, we obtain from (\ref{KLcasimir-maximal})
\begin{equation}
  (\epsilon_0)_{\rm SUSY}=-\frac{N(N^2-1)}{6r}\stackrel{N\rightarrow\infty}{\longrightarrow}
  -\frac{N^3}{6r}\ .
\end{equation}
From the interpretation in our previous paragraph, we think it is likely that
the disagreement of the two quantities is simply due to the fact that the gravity
dual and the 5d SYM computed different observables.\footnote{See, however, the reference
\cite{Minahan:2013jwa} for discussions on different possibilities of interpreting 
these results.} Assuming our interpretation, it will
be interesting to study what kind of computation should be done in the gravity side
to reproduce $(\epsilon_0)_{\rm SUSY}$. We think the key is to keep SUSY manifest in
the holographic renormalization computations, as this was what
yielded two different Abelian observables (\ref{KLcasimir-abelian}).

We also mention in passing that one can define a `supersymmetric version' of
Renyi entropy \cite{Nishioka:2013haa} in SUSY QFTs. This supersymmetric version
can be computed more easily in SUSY QFTs, similar to the supersymmetric version
of the Casimir energy that we explained here. We expect that there should be 
many supersymmetric observables of this sort.

Finally, let us consider (\ref{KLcasimir}) at more general points in the chemical potential
space. For instance, if one tries to repeat the computation of Casimir energies
from the free QFT consideration, clearly we have less symmetries which constrain
the regulator in the oscillator sum. So it might be that the quantity
could depend on the regularization scheme, and the localization computation
might have made an implicit assumption on it to get the result (\ref{KLcasimir}).
Since the maximal SUSY point (\ref{KLmaximal-SUSY}) appears to remove all such possible
ambiguities, we expect that any implicit assumptions that could have been made
in the path integral will not spoil (\ref{KLcasimir-maximal}). On the other hand,
even after turning on many chemical potentials, observables like `Casimir force'
that can be derived from the Casimir energy should be physical. So one should
be able to define both versions of Casimir energies $\epsilon_0$,
$(\epsilon_0)_{\rm SUSY}$ at most general values of chemical potentials.
It is not clear to us whether our computation captures such a physical quantity
at general value of chemical potential at all.

\subsection{The partition function on $\mathbb{CP}^2\times S^1$}

In this subsection, we discuss another expression of the superconformal index of
the $(2,0)$ theory of the same schematic form (\ref{KL6d-index-factorize}), which
this time takes a manifest index form. Following \cite{Kim:2013nva}, let us
explain this index for the $U(N)$ gauge group only. The index takes the following form:
\begin{eqnarray}\label{KLindex-CP2}
  \hspace*{-0.7cm}Z_{S^5\times S^1}(\beta,m,a_i)&=&\frac{1}{N!}\sum_{s_1,\cdots,s_N=-\infty}^\infty
  \oint\prod_{I=1}^Nd\lambda_Ie^{-S_0(\lambda,s,\beta)}Z_{\mathbb{R}^4\times T^2}
  \left(\frac{i\beta\omega_1}{2\pi},i\lambda-s\beta a_1,
  \omega_{21},\omega_{31},m-\frac{\omega_1}{2}\right)\nonumber\\
  \hspace*{-0.7cm}&&\hspace{-2cm}\cdot   Z_{\mathbb{R}^4\times T^2}
  \left(\frac{i\beta\omega_2}{2\pi},i\lambda-s\beta a_2,
  \omega_{32},\omega_{12},m-\frac{\omega_2}{2}\right)
  \cdot  Z_{\mathbb{R}^4\times T^2}
  \left(\frac{i\beta\omega_2}{2\pi},i\lambda-s\beta a_2,
  \omega_{32},\omega_{12},m-\frac{\omega_2}{2}\right), \nonumber\\
  S_0(\lambda,\beta)&=&\beta\sum_{I=1}^N\left(-\frac{s_I^2}{2}+is_I\lambda_I\right)\ , 
\end{eqnarray}
where again the index $Z_{\mathbb{R}^4\times T^2}(\tau,v,\beta,\epsilon_{1,2},m_0)$ is
used as building blocks, and we use the notation $\omega_i=1+a_i$,
$\omega_{ij}=\omega_i-\omega_j$. The `Coulomb VEV' parameter $\lambda$ is taken to be $\lambda_I\equiv\lambda^\prime_I-is\beta\zeta$ with any positive $\zeta$, where
$\lambda^\prime_I$ are variables whose integration contours are almost at the real axis
in the range $0\leq\lambda^\prime_I\leq 2\pi$. The precise integration contour will be
explained below, which goes around the poles in a specific manner. $\zeta$ appeared in
\cite{Kim:2013nva} as a freedom to choose the path integral contour for some fields,
and can be any number as long as it is positive. (The index will not depend on its value.
In \cite{Kim:2013nva}, it was parametrized by $\zeta=\frac{4}{\xi-1}$ with $\xi>0$.)

The contour for $\lambda_I^\prime$ was heuristically motivated in \cite{Kim:2013nva},
and was checked to yield reasonable results, but it is not rigorously derived so far.
The contour prescription is obtained as follows. We first take all $a_i$ parameters
and $m-\frac{1}{2}$ to be imaginary, in which case there will be many poles in the
$Z_{\mathbb{R}^4\times T^2}$ factors at the line ${\rm Im}   (\lambda_I)=0$. When
one considers an integral with all $s_I\neq s_J$, the integral contour is taken to be
along real $\lambda_I^\prime$ line between $1\leq\lambda^\prime_I\leq 2\pi$, which
does not hit any poles. When some $s_I$'s are equal, then the above real axis contour
will hit some poles on the real axis. In such a case, we slightly deform the contour,
or equivalently the poles away from the real axis, as follows. In each
$Z_{\mathbb{R}^4\times T^2}$ factor, the effective Omega parameters
$\epsilon_1,\epsilon_2$ are taken to be imaginary. The pole or contour deformation is
obtained by giving infinitesimal real shifts to these imaginary parameters as
\begin{equation}
  \epsilon_1+\varepsilon\ ,\ \ \epsilon_2+\varepsilon\ ,
\end{equation}
with $0<\varepsilon\ll 1$. This effectively deforms the contour to go around the poles
in a specific way. Although with some motivations about this rule presented in
\cite{Kim:2013nva}, we should stress that this is just a working prescription at the
moment of writing this review.

This formula was derived from a 5d SYM on $\mathbb{CP}^2\times\mathbb{R}$, which
was obtained by first considering 6d $(2,0)$ theory on a supersymmetric
$S^5/\mathbb{Z}_K\times\mathbb{R}$ orbifold \cite{Kim:2012tr,Kim:2013nva}. The
orbifold acts as follows. Considering the round $S^5$ as a Hopf fibration over
$\mathbb{CP}^2$, its metric is given by
\begin{equation}
  ds^2(S^5)=r^2\left[ds^2(\mathbb{CP}^2)+(dy+V)^2\right]\ ,
\end{equation}
where $y\sim y+2\pi$ and $V$ is related to the Kahler 2-form $J$ of $\mathbb{CP}^2$
by $J=\frac{1}{2}dV$. $S^5/\mathbb{Z}_K$ is obtained by modding out the fiber direction
by
\begin{equation}\label{KLZK-S5}
  y\sim y+\frac{2\pi}{K}\ .
\end{equation}
Our strategy is to first consider the regime with large $K$,
and obtain a 5d Yang-Mills theory on $\mathbb{CP}^2\times\mathbb{R}$ whose gauge coupling
is proportional to $\frac{1}{K}$. The coupling will be small for large $K$, or in the
energy scale $\frac{1}{r}\sim E\ll\frac{K}{r}$. Of course our eventual interest is the
case with $K=1$, in which case the 5d SYM is strongly coupled at all energy scale
$E\gtrsim\frac{1}{r}$. The expression(\ref{KLindex-CP2}) is obtained by studying this 5d Yang-Mills theory
on $\mathbb{CP}^2\times\mathbb{R}$ at strong coupling.

We would like to consider supersymmetric orbifold, as this would yield 5d supersymmetric
Yang-Mills theory which admits some exact computations. However, the above action leaves
none of the $(2,0)$ Killing spinors invariant \cite{Kim:2012tr}. To make a supersymmetric
$\mathbb{Z}_K$ orbifold, one can make a simultaneous rotation on the spatial angle
(\ref{KLZK-S5}) in $SO(6)$, and also on the internal $SO(5)$ R-symmetry. In \cite{Kim:2013nva}, an infinite family of rotations by $\frac{2\pi}{K}$ angle was considered with the rotation
generator
\begin{equation}\label{KLZK-charge}
  j_1+j_2+j_3+\frac{3}{2}(R_1+R_2)+n(R_1-R_2)\ ,
\end{equation}
where the $\frac{2\pi}{K}$ rotation with $j_1+j_2+j_3$ generates (\ref{KLZK-S5}).
To make the $2\pi$ rotation with this charge to be an identity, we should take
$n$ to be half an odd integer. By construction of the charge (\ref{KLZK-charge}),
this $\mathbb{Z}_K$ commutes with a pair of supercharges $Q=Q^{++}_{---}$ and
$S=S^{--}_{+++}$ that we used to define the superconformal index. At $K\neq 1$,
various 5d SYMs labeled by different $n$ will describe inequivalent systems, as 
$\mathbb{Z}_K$
orbifolds are all different. At the strong coupling point $K=1$, there is no orbifolding,
so different 5d SYMs on $\mathbb{CP}^2\times\mathbb{R}$ are expected to be all
equivalent at the quantum level. In particular, we expect SUSY enhancement to $OSp(8^\ast|4)$.

At $K=1$, the index (\ref{KLSC-index-(2,0)}) is computed from 5d SYM on
$\mathbb{CP}^2\times S^1$, with suitable twisted boundary conditions of fields on
$S^1$ by the chemical potentials \cite{Kim:2012tr,Kim:2013nva}. Although we expect
that all SYMs labeled by $n$ would yield equivalent results, 5d SYMs are very different 
at different values of $n$. The formula (\ref{KLindex-CP2}) is obtained from the SYM
associated with $n=-\frac{1}{2}$, which has a virtue of showing $8$ supercharges
explicitly in the 5d SYM action, including $Q,S$ above. The QFT at $n=-\frac{3}{2}$
was first discovered in \cite{Kim:2012tr}, and the perturbative index (without including
instantons) of this QFT at large $K$ was also computed.

The action and SUSY transformation of fields are explained in \cite{Kim:2013nva},
which we shall not repeat here. The derivation of the index (\ref{KLindex-CP2}) goes
in a similar way as that of the index (\ref{KLSC-index-1}) from the $S^5$ partition
function, which can be found in \cite{Kim:2013nva}. (Just like on $S^5$, the
`derivation' again assumes some localized nature of the instanton saddle point
configurations.) Here we would just like to stress a few qualitative
differences in the formula (\ref{KLindex-CP2}), compared to (\ref{KLSC-index-1}).

Firstly, the complex structure parameter $\tau_i=\frac{i\beta\omega_i}{2\pi}$
in each $Z_{\mathbb{R}^4\times T^2}$ provides the factor
$e^{2\pi i\tau_i k_i}=e^{-\beta\omega_i k_i}$ in the $k_i$ instanton coefficients.
So the expression(\ref{KLindex-CP2}) manifestly takes the form of an index at $\beta\gg 1$.
This should be clear since the 5d SYM is put on $\mathbb{CP}^2\times S^1$, explicitly
having a time direction, with $\beta$ being the circumference length of the circle. This is in
contrast to the $S^5$ partition function, where the
time direction had to `emerge' at strong coupling $\beta\gg 1$.

Secondly, apart from the three instanton summations in
$Z_{\mathbb{R}^4\times T^2}$ factors, the expression (\ref{KLindex-CP2}) has extra summations over
integers $s_1,\cdots,s_N$. This originates from a more complicated saddle point structure
of the path integral. To explain it, let us first fix our orientation convention on
$\mathbb{CP}^2$. We can decompose 2-forms $\mathbb{CP}^2$ to self-dual and anti-self-dual
2-forms. Our convention is such at the Kahler 2-form $J$ is in the anti-self-dual part.
Then, the saddle point first admits singular self-dual instantons, localized at the
fixed points of the $U(1)^2$ rotations generated by $j_1-j_2$, $j_2-j_3$. The summation over
these instanton numbers generate the series expansion with $e^{2\pi i\tau_i k_i}$ in
the three $Z_{\mathbb{R}^4\times T^2}$ factors. With nonzero $\lambda$ which breaks
$U(N)$ to $U(1)^N$, these self-dual instantons are all $U(1)^N$ instantons, just like
the instantons appearing in the Nekrasov partition functions.
On top of these, it turns out that one could also have anti-self-dual field configurations
$F^-=\frac{2s}{r^2}J$, where $s={\rm diag}(s_1,\cdots s_N)$ with integer eigenvalues.
The $-\beta\sum_I\frac{s_I^2}{2}$ term in $S_0$ is the contribution of these anti-self-dual
instantons to the energy $E$ in (\ref{KLSC-index-(2,0)}). These extra anti-self-dual
instantons play very nontrivial roles in making (\ref{KLSC-index-(2,0)}) to work.

One might worry that, summation over all the integers $s_I$ with unbound negative energy
weight $e^{\beta\sum_I\frac{s_I^2}{2}}$ will make the expression (\ref{KLindex-CP2}) divergent. However,
the $\lambda$ integration contour explained above will project this infinite sum over $s_I$
into a finite sum. The state which contributes with the most negative energy will be the
vacuum. (Here, by `energy' we mean $E-\frac{R_1+R_2}{2}$.)

In \cite{Kim:2013nva}, the expression (\ref{KLindex-CP2}) was used to study various aspects of the
$(2,0)$ theory. Here we shall explain two studies made there. Firstly, we shall
explain how the unrefined index of section 2.2 appears from this approach, as this
will illustrate how the formula (\ref{KLindex-CP2}) works in the simplest setting. Secondly, we shall
explain the systematic series expansion of the expression (\ref{KLindex-CP2}) in terms of fugacities
at some finite $N$ $(>1)$, keeping all independent chemical potentials generic. This
should probably be the strongest virtue of the expression (\ref{KLindex-CP2}).

We first consider the unrefined index at $m=\frac{1}{2}-a_3$.
From the expressions of $\epsilon_1$, $\epsilon_2$, $m_0$ appearing in the three factors
of the expression (\ref{KLindex-CP2}), one can show that they satisfy $m=\epsilon_-$ in the first
factor, $m=-\epsilon_-$ in the second factor, and $m=\epsilon_+$ in the third factor.
So the simplification pattern of the integrand is similar to the $S^5$ partition
function. Thus, one obtains \cite{Kim:2013nva}
\begin{equation}\label{KLCP2-unrefine}
  \frac{1}{\eta(\frac{i\beta\omega_3}{2\pi})^N}\cdot\frac{1}{N!}
  \oint[d\lambda_I]\sum_{s_1,\cdots,s_N=-\infty}^\infty
  e^{\frac{\beta}{2}\sum_Is_I^2-i\sum_I s_I\lambda_I}
  \prod_{I<J}2\sinh\frac{i\lambda_{IJ}-\beta s_{IJ}a_1}{2}\cdot
  2\sinh\frac{i\lambda_{IJ}-\beta s_{IJ}a_2}{2}\ ,
\end{equation}
where the first factor $\eta(\frac{i\beta}{2\pi})^{-N}$ comes from $Z_{\rm inst}$
from the third factor of the expression (\ref{KLindex-CP2}) at $m=\epsilon_+$.\footnote{We have included
the zero point energy factors of these instanton particles, to obtain $\eta^{-N}$.}
The integrand is so much simplified that there are no poles of $\lambda_I^\prime$
on their real axes. Thus, the contour integral can be taken along the real axes of
$\lambda^\prime_I$, or along the ${\rm Im} \lambda_I=-s_I\beta\zeta$ line. Consider
the complex variable $z_I=e^{-i\lambda_I}$. Since the only pole of the integrand appears
at $z_I=0$, one can continuously deform the integration contour to the unit circles
$|z_I|=1$ (namely ${\rm Im} \lambda_I=0$). The integral (\ref{KLCP2-unrefine}) can be
done easily \cite{Kim:2013nva}, which yields
\begin{equation}
  e^{\beta\omega_3\frac{N(N^2-1)}{6}}\eta(\frac{i\beta\omega_3}{2\pi})^{-N}
  \prod_{n=1}^{N-1}(1-e^{-n\beta\omega_3})^{N-n}=
  e^{\beta\omega_3\left(\frac{N(N^2-1)}{6}+\frac{N}{24}\right)}
  \prod_{s=0}^\infty\prod_{d=1}^N\frac{1}{1-e^{-\beta\omega_3(d+s)}}\ .
\end{equation}
The spectrum part of this index is exactly the same as (\ref{KLunrefined-spectrum}),
with $d=1,\cdots,N$ for $U(N)$ Casimir operators.

We next consider the vacuum energy factor. First of all, we go back to the integral
expression (\ref{KLCP2-unrefine}) and trace where the vacuum is coming from, among
the various saddle points of the 5d SYM. The vacuum comes from the configuration
in which $e^{\frac{\beta}{2}\sum_I s_I^2}$ factor in (\ref{KLCP2-unrefine}) is the largest. 
Apparently, taking all $s_I$'s to be arbitrary large, it might look that
one can make this factor as large as possible. If this were the case, then the index
(\ref{KLindex-CP2}) would not have made sense. But from the structure of
the contour integral (\ref{KLCP2-unrefine}), one cannot make $s_I$ to be arbitrary large.
This is due to the term $e^{-i\sum_Is_I\lambda_I}$, which is part of the classical action.
Physically, this term comes from a term of the form \cite{Kim:2012tr}
\begin{equation}\label{KLKCS}
  \int J\wedge {\rm tr}\left(A\wedge dA-\frac{2i}{3}A^3\right)
\end{equation}
in the action on $\mathbb{CP}^2\times S^1$ \cite{Kim:2012tr,Kim:2013nva}. Namely,
a magnetric flux $F\sim J$ induces an electric charge via the Kahler-Chern-Simons
term (\ref{KLKCS}). This induces a phase $e^{-i\sum_I s_I\lambda_I}$ in (\ref{KLCP2-unrefine}).
So for the contour integral to be nonzero, the rest of the measure in (\ref{KLCP2-unrefine})
should provide a phase which can cancel $e^{-i\sum_I s_I\lambda_I}$. Since the measure
consists of a product of $N^2-N$ sine functions, there are only finitely many values of
$s_1,\cdots,s_N$ for which the integral is nonzero. It turns out that the maximal value of
$e^{\frac{\beta}{2}\sum_I s_I^2}$ is obtained at
\begin{equation}
  (s_1,\cdots,s_N)=(N-1,N-3,\cdots,-(N-3),-(N-1))\ ,
\end{equation}
or any other configurations obtained by permuting the above $s_I$ fluxes. Summing over
$N!$ such fluxes, one obtains the following contribution
\begin{equation}
  \frac{1}{N!}\cdot N! e^{\frac{\beta}{2}((N-1)^2+(N-3)^2+\cdots+(-N+1)^2)}=
  e^{\beta\frac{N(N^2-1)}{6}}\ .
\end{equation}
Collecting the other factors coming from the sine functions, one finds extra
$e^{\frac{a_3\beta N(N^2-1)}{6}}$, and by expanding the instanton correction
$\eta(\frac{i\beta}{2\pi})^{-N}$, one obtains $e^{\frac{N\beta\omega_3}{24}}$ at lowest
energy. So combining all, one finds
$1\cdot\exp\left[\beta\omega_3\left(\frac{N(N^2-1)}{6}+\frac{N}{24}\right)\right]$.
This illustrates
that, from the viewpoint of SYM on $\mathbb{CP}^2\times\mathbb{R}$, the 6d CFT vacuum
and its energy $\sim N^3$ appear in a highly nontrivial manner, by `exciting' many
non-perturbative anti-self-dual instantons. It should be very interesting to understand
this vacuum structure more directly.

We find that the vacuum energy
\begin{equation}\label{KLCP2-casimir}
  (\epsilon_0)_{\rm SUSY}=-\omega_3\left(\frac{N(N^2-1)}{6}+\frac{N}{24}\right)
\end{equation}
is \textit{not} the same as (\ref{KLcasimir}) computed from the $S^5$ partition
function in general. This is not surprising because, as we explained in the previous
subsection, we have not too strong symmetry in general which could constrain the
regularization/renormalization of the path integral, so the two SYM computations on
$S^5$ and $\mathbb{CP}^2\times S^1$ could have implicitly chosen inequivalent
regularization schemes. It is not clear to us at the moment if any of the two is
physically meaningful. However, when $m=\frac{1}{2}$ and $a_i=0$, recall that we have
maximal SUSY $SU(4|2)$ which we expect to constrain the regularization completely.
Indeed, at this point ($\omega_i=1$), the two results (\ref{KLcasimir}) and
(\ref{KLCP2-casimir}) agrees with each other, supporting our expectation.

The final subject of this subsection is the general
index with all four chemical potentials turned on. The full expression (\ref{KLindex-CP2})
is too complicated for us to handle exactly, but now we can systematically make a
low energy fugacity expansion. This has been done in \cite{Kim:2013nva} for various
values of $N$ until a few low orders in $e^{-\beta}$.

Let us define $q\equiv e^{-\beta}$, $y\equiv e^{\beta(m-\frac{1}{2})}$,
$y_i\equiv e^{-\beta a_i}$ (satisfying $y_1y_2y_3=1$). We shall be expanding the index
by assuming $q\ll 1$, keeping $y,y_i$ to be of order $1$.
Firstly, for general $N$, the expression (\ref{KLindex-CP2}) was computed up to
$\mathcal{O}(q^2)$. The result apart from the zero point energy factor is
\begin{eqnarray}\label{KLCP2-expand}
  Z_{S^5\times S^1}&=&1+qy
  +q^2\left[2y^2+y(y_1+y_2+y_3)-(y_1^{-1}+y_2^{-1}+y_3^{-1})+y^{-1}\right]+\mathcal{O}(q^3)
\end{eqnarray}
for $N\geq 2$. (The exact index at $N=1$ was worked out in \cite{Kim:2013nva} separately.)
The result is independent of $N$ for $N\geq 2$. In fact, empirically in all studies done
in \cite{Kim:2013nva}, the index will turn out to be independent of $N$ at
$q^k$ order if $k\leq N$. This is a natural thing to expect for a CFT with large $N$
gravity dual. This is because $E\ll N$ is the regime in which supergravity approximation
of the string/M-theory is valid, and the gravity spectrum is independent of $N$. Of course,
the $N$ independence of the spectrum up to the threshold $E=N$ is too much to expect, but
it often happens at least in the BPS sector that $E\sim N$ is the threshold beyond which
the `stringy exclusion' behaviors \cite{McGreevy:2000cw} start to appear. 
The $N$ independent index (\ref{KLCP2-expand}) completely agrees with the large $N$ 
supergravity index on $AdS_7\times S^4$, which is a consistency check of the expression (\ref{KLindex-CP2}).

As explained in \cite{Kim:2013nva}, the analysis at higher orders in $q$ becomes quickly
complicated, due to the appearance of many instanton saddle points contributing to the formula
(\ref{KLindex-CP2}). The studies are made for $N=2,3$ till $q^3$ order in
\cite{Kim:2013nva}. After adding many contributions from various saddle points, each
of them acquiring contributions from many residues in the contour integral, the $q^3$
order corrections to (\ref{KLCP2-expand}) for $N=2,3$ are given by
\begin{eqnarray}
  U(2)&:&q^3\left[2y^3+2y^2(y_1+y_2+y_3)+y\left(y_1^2+y_2^2+y_3^2-\frac{1}{y_1}-\frac{1}{y_2}
  -\frac{1}{y_3}\right)\right.\nonumber\\
  &&\hspace{2cm}\left.-\left(\frac{y_1}{y_2}+\frac{y_2}{y_1}+\frac{y_2}{y_3}+\frac{y_3}{y_2}
  +\frac{y_3}{y_1}+\frac{y_1}{y_3}\right)+y^{-1}(y_1+y_2+y_3)\right]\label{KLU(2)}\ , \\
  U(3)&:&q^3\left[3y^3+2y^2(y_1+y_2+y_3)+y\left(y_1^2+y_2^2+y_3^2-\frac{1}{y_1}-\frac{1}{y_2}
  -\frac{1}{y_3}\right)\right. \nonumber\\
  &&\hspace{2cm}\left.-\left(\frac{y_1}{y_2}+\frac{y_2}{y_1}+\frac{y_2}{y_3}+\frac{y_3}{y_2}
  +\frac{y_3}{y_1}+\frac{y_1}{y_3}\right)+y^{-1}(y_1+y_2+y_3)\right]\ .\label{KLU(3)}
\end{eqnarray}
The $U(3)$ result (\ref{KLU(3)}) completely agrees with the large $N$ supergravity index
on $AdS_7\times S^4$, presumably because $k=N$ is the threshold until which the BPS
spectrum is independent of $N$. For $U(2)$, we see from (\ref{KLU(2)}) that one
state is missing compared to the large $N$ index, i.e. $2y^3$ vs. $3y^3$ in the first
terms. It will be interesting to study the $U(2)$ index at very high order in $q$,
and investigate a truly unexplored sector of the 6d $(2,0)$ theory beyond supergravity.

\section{Discussions}

In this review, we explained the recent progress on the 6d SCFT partition functions
in the Coulomb and the symmetric phases, focusing on the Coulomb branch indices in the
Omega background and the superconformal index on $S^5\times S^1$. The two observables
are closely related, and we explained their relations and the physics contained in
these indices with the example of $(2,0)$ theory. In this section, we discuss some
open problems, and some recent progress on this subject that we could not properly
review in this work.

As we tried to emphasize in section 2, the computation of the indices in
the Coulomb phase does not really rely on the 5d SYM description. We can rather
understand it as a direct string theory computation, in the very background which
is used to engineer the 6d SCFT itself. The references \cite{Hwang:2014uwa,Kim:2014dza} discuss more
subtle $(1,0)$ theories in the Coulomb phase in a similar manner, such as those living on the
M5-M9-brane system. \cite{Hwang:2014uwa} also explain how one can extract out the 
6d Coulomb branch partition functions
from string theory computations. So we claim that the expressions like
(\ref{KLSC-index-1}) and (\ref{KLindex-CP2}) for the 6d superconformal indices should be
using such `intrinsic' partition functions on $\mathbb{R}^4\times T^2$, which are
defined and computed without referring to the 5d SYM. However, it seems (at least so far)
that we have no way to even motivate the curved space partition functions results like
(\ref{KLSC-index-1}) and (\ref{KLindex-CP2}), set aside derivations, without using the
the 5d SYM descriptions. Of course one can hope these formulae to be true even when
small circle reductions of the 6d CFT does not flow to weakly coupled 5d SYM, as all
the ingredients appearing in these formulae can be addressed without referring to 5d SYM.
This makes us  suspect that there should be a more abstract way of understanding these
formulae, perhaps directly using string theory. However, we do not know if we can realize
$S^5\times\mathbb{R}$ background and put 6d CFT there directly in the string theory
setting.

We have presented two different expressions (\ref{KLSC-index-1}) and 
(\ref{KLindex-CP2}) for the superconformal indices for a given theory, and found the same
physics in various sectors of the $(2,0)$ theory when we could make concrete studies of
them. Of course more basic question is whether the two partition functions are identically
the same, perhaps modulo the Casimir energy factors which might be ambiguous in general.
Answering this question would have to do with making a strong-coupling re-expansion of
the ingredients $Z_{\mathbb{R}^4\times T^2}$ in (\ref{KLSC-index-1}).

Related to the last question, and also for applying our findings to more general
$(1,0)$ SCFTs, it would be very important to understand the modular properties of
the partition functions  (\ref{KL2d-expansion}) and (\ref{KL1d-expansion}) better. Knowing its
modular property under $\tau\rightarrow-\frac{1}{\tau}$ means that we can make a
strong coupling expansion of (\ref{KLSC-index-1}). At this point, we should emphasize
the studies of the elliptic genus which appears as the coefficients of (\ref{KL2d-expansion})
for various self-dual strings. Namely, assuming large Coulomb VEV $v$,
$Z_{n_I}(\tau,\epsilon_{1,2},m)$ were computed from various 2d gauge theories living
on the self-dual strings. It has been first studied for the $A_{N-1}$ type $(2,0)$ strings
in \cite{Haghighat:2013gba,Haghighat:2013tka}, which are called M-strings. More interesting
$(1,0)$ self-dual strings have been studied this way. For instance, the $(1,0)$ strings for
M2-branes suspended between M5-M9 branes are called E-strings, whose elliptic genera were
systematically computed from 2d QFT \cite{Kim:2014dza}. Some other strings for $(1,0)$
theory engineered by F-theory were studied recently in \cite{Haghighat:2014vxa}.
With these elliptic genera known, it will be in principle possible to trace how to
make a strong coupling re-expansion of the integrand of (\ref{KLSC-index-1}), and
address the index for a variety of $(1,0)$ CFTs.\footnote{Note that the studies of 
this paper either used special properties of the Abelian theory,
or relied on SUSY enhancement which are not available for $(1,0)$ theories. Also, 
the $\mathbb{CP}^2\times S^1$ index might have used too many ingredients of the 
$(2,0)$ theory, so it is not clear whether the strategy will go through 
well for all $(1,0)$ CFTs.}

Even if one forgets about the application to the symmetric phase observables,
computing the elliptic genera of various $(1,0)$ self-dual strings would be
very valuable by itself. And this is quite challenging in general, as engineering
a weakly coupled 2d gauge theory on the worldsheet is not always easy. For instance,
the task becomes relatively easier if the self-dual strings can be engineered using
D-branes subject to various boundary conditions \cite{Haghighat:2013gba,Haghighat:2013tka,Kim:2014dza,Haghighat:2014vxa}. However,
many interesting 6d CFTs are engineered from F-theory, which involves exotic
7-branes. It will be interesting to see how much we can learn about them 
from 2d gauge theories.

Finally, it should be interesting to explore the 6d CFT partition functions on other
curved manifolds, presumably using various 5d SYM approaches. In this paper we tried not to
mention 5d SYM description when unnecessary, e.g. in section 2, for the sake of consistency
and also for logical clarity. But of course 5d SYM provides extremely useful viewpoint to
study this system. In the very limited class of SUSY observables that we studied, 
the only subtlety of 5d SYM that we could find was the small instanton issue.
So we would very much like to know how much 5d SYM can be teaching us about higher
dimensional CFTs \cite{Douglas:2010iu}.

\vskip 0.5cm

\hspace*{-0.8cm} {\bf\large Acknowledgements}
\vskip 0.2cm

\hspace*{-0.75cm} We would like to thank Chiung Hwang, Hee-Cheol Kim, Joonho Kim, Sung-Soo Kim, Eunkyung Koh, Sungjay Lee, Jaemo Park, Cumrun Vafa for their collaborations
with the authors, on the materials covered in this paper. We would also like to thank Dario Martelli, Shiraz Minwalla and especially Hee-Cheol Kim for helpful discussions on 
the related subjects while we were writing this review. We also like to thank Yuji Tachikawa for careful reading of this manuscript and many useful comments.   This work is supported in part 
by the National Research Foundation of Korea (NRF) Grant  No. 2012R1A1A2042474 (SK), 2012R1A2A2A02046739 (SK), 2015R1A2A2A01003124 (SK), 2006-0093850 (KL).

\section{Appendix. Off-shell supergravity analysis on $S^5$}

By making the KK reduction from $S^5\times S^1$ with twists by $a_i$,
our background fields are given by
\begin{equation}
  C=i\alpha^2 a_in_i^2d\phi_i\ ,\ \ \alpha^2=\frac{1}{1-a_i^2n_i^2}\ .
\end{equation}
The field $C$ is imaginary. By comparing the gravitino SUSY condition of
\cite{Kugo:2000af} and the 5d reduction of the Killing spinor equation for the two
spinors on our background, one finds that $v_{\mu\nu}=-\frac{i}{4\alpha}(dC)_{\mu\nu}$ for
the antisymmeric field in the Weyl multiplet, $b_\mu=0$ for the dilatation gauge field,
and $V_\mu=-C_\mu\frac{\sigma_3}{2}$ for the $SU(2)_R$ gauge field.
The Killing spinor equation is
\begin{equation}
  D_a\epsilon=\left[\frac{i}{8\alpha}\gamma_{abc}(dC)^{bc}+i\gamma_a
  \left(-\alpha\frac{\sigma_3}{2}-\frac{1}{4\alpha}\gamma_{bc}V^{bc}+
  \frac{i}{2\alpha}\gamma^b\nabla_b\alpha\right)\right]\epsilon\ ,
\end{equation}
where $D_a\epsilon=\left(\nabla_a+C_a\frac{\sigma_3}{2}\right)\epsilon$. $a,b,c,\cdots$
are frame indices. The conjugate spinor $\epsilon^\dag$ is literally taken to be the Hermitian
conjugate in our Euclidean theory, so it satisfies
\begin{equation}
  D_a\epsilon^\dag=\epsilon^\dag\left[-\frac{i}{8\alpha}\gamma_{abc}V^{bc}-i
  \left(-\alpha\frac{\sigma_3}{2}-\frac{1}{4\alpha}\gamma_{bc}V^{bc}-
  \frac{i}{2\alpha}\gamma^b\nabla_b\alpha\right)\gamma_a\right]\ ,
\end{equation}
where $D_a\epsilon^\dag=\nabla_a\epsilon^\dag-\epsilon^\dag C_a\frac{\sigma_3}{2}$.
The imaginary nature of $C,V$ is all taken into account.
Below we shall study some bosonic equations which are derived from the above
Killing spinor equation, from which we determine various geometric quantities. As we
are physically quite confident from 6d arguments that (\ref{KLbackground}) should be a SUSY 
background, we shall only study a subset of the bosonic equations to determine the fields, 
rather than completely solving them.

We study the differential conditions satisfied by the spinor bilinears.
One first obtains
\begin{equation}\label{KLscalar}
  \nabla_a\left(\alpha\ \epsilon^\dag\epsilon\right)=-i\xi^b(dC)_{ab}\ .
\end{equation}
where $\xi^a\equiv\epsilon^\dag\gamma^a\epsilon$.
One similarly obtains the following condition for the vector bilinear:
\begin{equation}
  \nabla_b\xi_a=-\frac{i}{4\alpha}(\epsilon^\dag\gamma_{abcd}\epsilon)(dC)^{cd}
  +i\alpha(\epsilon^\dag\sigma_3\gamma_{ab}\epsilon)+\frac{i}{\alpha}
  (\epsilon^\dag\epsilon)(dC)_{ab}-\frac{2}{\alpha}\xi_{[a}\nabla_{b]}\alpha
  -\frac{1}{\alpha}\delta_{ab}\xi^c\nabla_c\alpha\ .
\end{equation}
Note that all but the last term is antisymmetric in $a,b$. So one obtains
\begin{equation}\label{KLvector1}
  \nabla_a\xi_b+\nabla_b\xi_a=-2g_{ab}\ \frac{\xi\cdot\nabla\alpha}{\alpha}
\end{equation}
and
\begin{equation}\label{KLvector2}
  d\xi=\frac{i}{4\alpha}\star(V\wedge \xi)
  -\frac{i}{\alpha}fV+\frac{2}{\alpha}\xi\wedge d\alpha+2\alpha X^3\ ,
\end{equation}
where $f=\epsilon^\dag\epsilon$, $V_{\mu\nu}\equiv(dC)_{\mu\nu}$,
$X^3_{\mu\nu}=-\frac{i}{2}\epsilon^\dag\sigma^3\gamma_{\mu\nu}\epsilon$.
We shall need the expression for $X^3$ from (\ref{KLvector2}) later.
There are more differential conditions for the tensor bilinears. We will
not need to consider them.
We also study the algebraic conditions satisfied by the bilinears. 
In our Euclidean theory, the algebraic conditions become
\begin{equation}\label{KLalgebraic}
  \xi^\mu\xi_\mu=f^2\ ,\ \ i_\xi X^3=0\ ,\ \ i_\xi \star X^3=-fX^3\ ,\ \
  4(X^3\cdot X^3)_{\mu\nu}=-f^2g_{\mu\nu}+\xi_\mu\xi_\nu
\end{equation}
etc. We shall not consider other 2-form bilinears $X^1,X^2$ in this paper.

A possible guess for  $\xi=\xi^\mu\partial_\mu$ is the following. A highly well-motivated
conjecture for $\xi$ is that it should generate the bosonic symmetry
for $\mathcal{Q}^2$ algebra, where $\mathcal{Q}=Q+S$ is the supercharge associated
with our index. So we try
\begin{equation}\label{KLvector-guess}
  \xi=\sum_{i=1}^3\omega_i\partial_{\phi_i}\ .
\end{equation}
Firstly, this trivially solves (\ref{KLvector1}), since $\xi$ is
a Killing vector which leaves $\alpha$ invariant. Then we plug this $\xi$ into
the right hand side of (\ref{KLscalar}). Here we can nontrivially test our educated guess
(\ref{KLvector-guess}), since $-i\xi^b(dC)_{ab}$ is integrable with the above $\xi$.
The solution to (\ref{KLscalar}) with $f=\epsilon^\dagger\epsilon$ is
\begin{equation}\label{KLf}
  f=\frac{1+a_in_i^2}{\sqrt{1-a_i^2n_i^2}}\ .
\end{equation}
One can check that this result is also compatible with the algebraic conditions 
(\ref{KLalgebraic}).

The only background fields of \cite{Kugo:2000af} that we have not determined yet are
${\rm D}$ in the Weyl multiplet, and the background vector multiplet fields. ${\rm D}$ can
be determined in the above background by studying the Weyl multiplet gaugino
SUSY variation. Contracting the SUSY condition with $\epsilon^\dag$,
one obtains
\begin{equation}
  f{\rm D}=\frac{i}{2\alpha}
  \xi_a\left(\nabla_bV^{ba}-\frac{i}{4\alpha}\epsilon^{abcde}
  V_{bc}V_{de}+\frac{\nabla_b\alpha}{\alpha}V^{ba}\right)
  -2iX^3_{ab}V^{ab}-\frac{f}{2\alpha^2}V^{ab}V_{ab}\ .
\end{equation}
Inserting (\ref{KLvector2}) for $X^3$, and plugging in $g_{\mu\nu}$, $V=dC$, $\alpha$
of (\ref{KLbackground}), $\xi$ of (\ref{KLvector-guess}), $f$ of (\ref{KLf}), one finds
a big simplification, after which ${\rm D}$ is simply given by
\begin{equation}
  {\rm D}=2(a_1^2+a_2^2+a_3^2)\alpha^2\ .
\end{equation}

We also need to determine the background vector field of various sorts. There are
various flavor background gauge fields, coupling to the hypermultiplet, and also
one auxiliary vector multiplet whose scalar VEV should provide the Yang-Mills coupling.
In the notation of \cite{Kugo:2000af}, SUSY condition for gauge multiplet gaugino
is given by
\begin{eqnarray}
  \delta\chi^A&=&\frac{i}{2}F_{\mu\nu}\gamma^{\mu\nu}\epsilon^A
  +\gamma^\mu D_\mu\phi\epsilon^A-D^A_{\ B}\epsilon^B+2\phi\eta^A\\
  &=&\frac{i}{2}(F_{\mu\nu}-\alpha^{-1}\phi V_{\mu\nu})\gamma^{\mu\nu}\epsilon^A
  +\alpha D_\mu(\alpha^{-1}\phi)\gamma^\mu\epsilon^A
  +(i\alpha\phi\sigma_3-D)^A_{\ B}\epsilon^B\nonumber
\end{eqnarray}
in our notation and normalization for fields, with 
\begin{equation}
  \eta_\pm=i\left(\alpha\sigma_3/2-\frac{1}{4\alpha}V_{\mu\nu}\gamma^{\mu\nu}
  +\frac{i}{2\alpha}\nabla_\mu\alpha\gamma^\mu\right)\epsilon_\pm\ .
\end{equation}
The $\pm$ signs are for the $SU(2)_R$ doublet, containing $R$ or $\frac{R_1+R_2}{2}$ 
as the Cartan.
The supersymmetric configurations, apart from possible singular behaviors, are
\begin{equation}
  F_{\mu\nu}=\alpha^{-1}\phi(dC)_{\mu\nu}\ ,\ \ D_\mu(\alpha^{-1}\phi)=0\ ,\ \
  D=i\alpha\phi\sigma_3\ .
\end{equation}
This is solved by $\phi=\alpha\phi_0$ with a constant
$\phi_0$, and $A_\mu=\phi_0C_\mu$, $D=i\alpha^2\phi_0\sigma_3$. For various background
vector multiplet fields, this will be enough. However, these configurations are also
legitimate saddle point configurations for the dynamical vector multiplet fields
in the path integral. There it will be necessary to include singular configurations
to the above solutions. There is only one background vector multiplet field appearing
in the vector multiplet action, whose nonzero scalar VEV sets   the Yang-Mills
coupling scale.
Namely, we take $(A^I_\mu,\chi^{IA},\phi^I)$ with $I=0,1,\cdots,n_V$, where $n_V$ is
the number of matter vector multiplet fields. In our case, $n_V=|G|$. 
There is one auxiliary scalar $\phi^0$, and the remaining $n_V$ scalars are 
arranged into a matrix $\phi$. The matter-gravity
coupling action is given by the cubic function $\mathcal{N}=C_{IJK}\phi^I\phi^J\phi^K$,
which we take as $\mathcal{N}=\phi^0{\rm tr}(\phi\phi)$ in our case. The background
fields are given by $A^0_\mu=C_\mu$, $\phi^0=\alpha$, $D^0=i\alpha^2\sigma_3$. So
these background fields are the `gravi-photon/dilaton' background. The vector
multiplet action can be read off from \cite{Kugo:2000af}. For instance, the bosonic
part of the vector multiplet action is given by
\begin{eqnarray}\label{KLvector-action}
  2g_{YM}^2e^{-1}\mathcal{L}_V&=&\left(-\frac{1}{2}D+\frac{1}{4}R+\frac{3}{16\alpha^2}V^2\right)
  \alpha\phi^2-\frac{1}{2\alpha}\phi^2V^2
  +\phi\left(2\partial^a\alpha D_a\phi-\frac{i}{4}\alpha^2(\sigma_3)_{AB}D^{AB}\right)
  \nonumber\\
  &&+\frac{\alpha}{2}F_{ab}F^{ab}+\alpha D^a\phi D_a\phi+\frac{\alpha}{2}D_{AB}D^{AB}
  +e^{-1}\frac{i}{4}\epsilon^{\mu\nu\rho\sigma\tau}C_\mu F_{\nu\rho}F_{\sigma\tau}\ ,
\end{eqnarray}
where trace is assumed. The first order term $2\phi\partial^a\alpha D_a\phi$ on 
the first line can be integrated by part, to yield the mass term $-(\partial^2\alpha)\phi^2$.

Now we consider the special SUSY configuration $A_\mu=\phi_0 C_\mu$,
$D=i\alpha^2\phi_0\sigma_3$, $\phi=\alpha\phi_0$ for \textit{dynamical}
vector multiplet fields. Plugging this into the action (\ref{KLvector-action}),
the saddle point action is given by
\begin{equation}
  S_0=\frac{4\pi^3{\rm tr}(\phi_0^2)}{g_{YM}^2\omega_1\omega_2\omega_3}\ .
\end{equation}
Restoring $r$ and making all parameters dimensionless $\beta=\frac{g_{YM}^2}{2\pi r}$,
$r\phi_0=\phi_{\rm new}$, one obtains
\begin{equation}
  S_0=\frac{2\pi^2{\rm tr}(\phi^2)}{\beta\omega_1\omega_2\omega_3}\ ,
\end{equation}
which is the classical measure used in (\ref{KLSC-index-1}). When singular self-dual
instanton strings are put on $S^5$ along $S^1$ at $(n_1,n_2,n_3)=(1,0,0)$, $(0,1,0)$,
$(0,0,1)$, as mentioned in section 3.1, there is an extra contribution to the classical
action. Supposing that $k_1,k_2,k_3$ self-dual instanton strings are put at three
circles in $U(1)^r\subset G_r$ part of the gauge group, the first and last term of
the second line of (\ref{KLvector-action}) makes additional contribution to the action.
The net action is \cite{Kim:2012qf}
\begin{equation}
  S_0=\frac{2\pi^2{\rm tr}(\phi^2)}{\beta\omega_1\omega_2\omega_3}+
  \sum_{i=1}^3\frac{4\pi^2k_i}{\beta\omega_i}\ .
\end{equation}

For the charged hypermultiplets, one should rely on a bit brute-force method of
constructing the SUSY action and transformation. One can follow \cite{Hosomichi:2012ek},
which constructs the action with one off-shell supersymmetry. For the hypermultiplet
with scalar $q_A$, complex fermion $\psi$, we introduce two complex auxiliary fields
$F_{A^\prime}$, following \cite{Hosomichi:2012ek}. The supersymmetric action (also
coupling with vector multiplet fields) is given by
\begin{eqnarray}
  \mathcal{L}_H&=&|D_\mu q^A|^2+|[\phi,q^A]|^2-\bar{q}_A(\sigma^I)^A_{\ B}[D^I,q^B]
  +\left(4-\alpha^2/4\right)|q^A|^2-\bar{F}_{A^\prime}F^{A^\prime}\\
  &&+i\psi^\dag\gamma^\mu D_\mu\psi+i\psi^\dag[\phi,\psi]+\sqrt{2}i\psi^\dag
  [\chi_A,q^A]-\sqrt{2}i[\bar{q}_A,\chi^{\dag A}]\psi-\frac{1}{8\alpha}\psi^\dag
  V_{ab}\gamma^{ab}\psi+\frac{i}{2\alpha}\partial_a\alpha\psi^\dag\gamma^a\psi\ .
  \nonumber
\end{eqnarray}
We presented the result for adjoint hypermultiplet, but the action for other
representations should also be clear. The action is invariant under the SUSY
transformation
\begin{eqnarray}
  \delta q^A&=&\sqrt{2}i\epsilon^{\dag A}\psi\ ,\ \
  \delta\bar{q}_A=\sqrt{2}i\psi^\dag\epsilon_A\ , \\
  \delta\psi&=&\sqrt{2}\left[-D_\mu q_A\gamma^\mu\epsilon^A+[\phi,q_A]\epsilon^A
  +\frac{3i}{2}\alpha q_A(\sigma^3)^A_{\ B}\epsilon^B
  +(\frac{i}{2\alpha}V_{ab}\gamma^{ab}+\frac{2}{\alpha}\partial_a
  \alpha \gamma^a)q_A\epsilon^A-iF_{A^\prime}\hat\epsilon^{A^\prime}\right]\ , \nonumber\\
  \delta F^{A^\prime}&=&\sqrt{2}\hat\epsilon^{\dag A^\prime}
  \left[-\gamma^\mu \nabla_\mu\psi-[\phi,\psi]+\frac{i}{8\alpha}V_{ab}\gamma^{ab}\psi
  +\frac{1}{2\alpha}\partial_a\alpha\gamma^a\psi\right]\ ,\nonumber
\end{eqnarray}
and the spinor $\hat\epsilon$ satisfies \cite{Hosomichi:2012ek}
\begin{equation}
  \epsilon^\dag\epsilon=\hat\epsilon^\dag\hat\epsilon\ ,\ \
  (\epsilon^A)^TC\hat\epsilon^{B^\prime}=0\ ,\ \
  \epsilon^\dag\gamma^\mu\epsilon+\hat\epsilon^\dag\gamma^\mu\hat\epsilon=0\ .
\end{equation}
To turn on the hypermultiplet mass $m$, we introduce one more background vector multiplet
for the hypermultiplet flavor symmetry, and give them supersymmetric background
values with $\phi^0\sim m$. Then one can couple this background vector field with the
above hypermultiplet in the same way as the dyanmical vector fields couple to the
hypermultiplets above.

\documentfinishBBL